\documentclass[twocolumn]{aastex62}

\usepackage{natbib}

\shorttitle{MANOS spectroscopic results}
\shortauthors{Devog\`{e}le et al.}

\begin{document}

\title{Visible spectroscopy from the Mission Accessible Near-Earth Object Survey (MANOS): Taxonomic dependence on asteroid size }

\correspondingauthor{Maxime Devog\`{e}le}
\email{mdevogele@lowell.edu}

\author[0000-0002-6509-6360]{Maxime Devog\`{e}le}
\affil{Lowell Observatory, 1400 W Mars Hill RD, Flagstaff, AZ 86001, USA}

\author[0000-0001-6765-6336]{Nicholas Moskovitz}
\affil{Lowell Observatory, 1400 W Mars Hill RD, Flagstaff, AZ 86001, USA}

\author[0000-0002-1506-4248]{Audrey Thirouin}
\affil{Lowell Observatory, 1400 W Mars Hill RD, Flagstaff, AZ 86001, USA}

\author{Annika Gustaffson}
\affil{Department of Physics and Astronomy, Northern Arizona University, Flagstaff, AZ 86011, USA
}

\author{Mitchell Magnuson}
\affil{Department of Physics and Astronomy, Northern Arizona University, Flagstaff, AZ 86011, USA
}

\author{Cristina Thomas}
\affil{Department of Physics and Astronomy, Northern Arizona University, Flagstaff, AZ 86011, USA
}

\author{Mark Willman}
\affil{University of Hawaii, Kalaoa, HI, USA\\
}

\author{Eric Christensen}
\affil{University of Arizona, Department of Planetary Sciences, Lunar and Planetary Lab, Tucson, AZ 85721, USA
}

\author{Michael Person}
\affil{Department of Earth, Atmospheric, and Planetary Sciences, Massachusetts Institute of Technology
77 Massachusetts Avenue, Cambridge, MA 02139 USA}

\author{Richard Binzel}
\affil{Department of Earth, Atmospheric, and Planetary Sciences, Massachusetts Institute of Technology
77 Massachusetts Avenue, Cambridge, MA 02139 USA}

\author{David Polishook}
\affil{Faculty of Physics, Weizmann Institute of Science, Rehovot 7610001, Israel\\
}

\author{Francesca DeMeo}
\affil{Department of Earth, Atmospheric, and Planetary Sciences, Massachusetts Institute of Technology
77 Massachusetts Avenue, Cambridge, MA 02139 USA}

\author{Mary Hinkle}
\affil{Department of Physics and Astronomy, Northern Arizona University, Flagstaff, AZ 86011, USA
}

\author{David Trilling}
\affil{Department of Physics and Astronomy, Northern Arizona University, Flagstaff, AZ 86011, USA
}

\author{Michael Mommert}
\affil{Lowell Observatory, 1400 W Mars Hill RD, Flagstaff, AZ 86001, USA}

\author{Brian Burt}
\affil{Lowell Observatory, 1400 W Mars Hill RD, Flagstaff, AZ 86001, USA}

\author{Brian Skiff}
\affil{Lowell Observatory, 1400 W Mars Hill RD, Flagstaff, AZ 86001, USA}

\begin{abstract}
The Mission Accessible Near-Earth Object Survey (MANOS) aims to observe and characterize small (mean absolute magnitude H $\sim$ 25 mag) Near-Earth Objects (NEOs) that are accessible by spacecraft (mean $\Delta v \sim 5.7$ km/s) 
and that make close approaches with the Earth (mean Minimum Orbital Intersection Distance MOID $\sim$ $0.03$ AU). We present here the first results of the MANOS visible spectroscopic survey. The spectra were obtained from August 2013 to March 2018 at Lowell Observatory's Discovery Channel 4.3 meter telescope, and both Gemini North and South facilities. In total, 210 NEOs have been observed and taxonomically classified. Our taxonomic distribution shows significant variations with respect to surveys of larger objects. We suspect these to be due to a dependence of Main Belt source regions on object size. Compared to previous surveys of larger objects \citep{Binzel_2019,Binzel_2004, Perna_2018}, we report a lower fraction of S+Q-complex asteroids of $43.8 \pm 4.6\%$. We associate this decrease with a lack of Phocaea family members at very small size. We also report higher fractions of X-complex and A-type asteroids of $23.8 \pm 3.3\%$ and $3.8 \pm 1.3 \%$ respectively due to an increase of Hungaria family objects at small size. We find a strong correlation between the Q/S ratio and perihelion distance. We suggest this correlation is due to planetary close encounters with Venus playing a major role in turning asteroids from S to Q-type. This hypothesis is supported by a similar correlation between the Q/S ratio and Venus MOID. 
\end{abstract}


\section{Introduction} \label{sec:intro}

Near-Earth Objects (NEOs) are defined by a perihelion distance $q < 1.3$ AU. The study of NEOs provides access to objects up to 3 orders of magnitude smaller than the smallest observable Main Belt asteroids (MBAs), where most NEOs are thought to have originated. NEOs are also the most accessible objects to spacecraft in the Solar System, enabling detailed study of their physical properties. Since the discovery of (433)~Eros in 1898, the number of known NEOs has continuously grown and now reaches over 20,000 objects as of April 2019. 

To date, a representative census of NEO physical properties exists only for the largest objects (equivalent diameter $D>1$ km). They have been studied using various techniques such as time-series photometry \citep[e.g.][]{Krugly_2002, Chang_2015}, spectrophotometry \citep[e.g.][]{Mommert_2016,Erasmus_2017, Navarro_2019}, spectroscopy \citep[e.g][]{Binzel_2004, Binzel_2019}, radar techniques \citep[e.g.][]{Ostro_2006}, and polarimetry \citep[e.g.][]{DeLuise_2007,Devogele_2018, Cellino_2018}. However, an equivalent census for sub-km NEOs, which represent more than 95\% of the currently known population, does not exist. The goal of this work and the Mission Accessible Near-Earth Object Survey (MANOS) is to address this issue.

MANOS is an observational survey of small (mean H $\sim$ 25 mag), mission accessible (mean $\Delta v \sim 5.7$ km/s) NEOs which experience close approaches to Earth (mean Earth Minimum Orbit Intersection Distance or MOID $\sim$ $0.03$ AU). $\Delta v$ in this context is defined as the impulse needed for a spacecraft to maneuver from low Earth orbit to a rendezvous with the asteroid in its orbit. It can be computed for NEOs using the approximation described by \citet{Shoemaker_1978}. 
MANOS provides comprehensive characterization of these objects by performing astrometic, photometric \citep{Thirouin_2016, Thirouin_2018}, and spectroscopic (this work) observations. The first observations started in late 2013 and the project is currently funded by the NASA Solar System Observations program through mid-2020.

The study of small NEOs is of importance for several reasons. It is currently estimated that there are $\sim 10^{7}$ objects with $D>10$ m, whereas $\sim 10^{4}$ have $D>100$ m \citep{Harris_2015,Trilling_2017}. The increasing numbers at small sizes implies higher probability of a small NEO impacting the Earth on relatively short ($<$decadal) timescales. To date, only three asteroids (2008 TC3, 2014 AA, and 2018 LA) have been telescopically observed prior to impact and all are smaller than 10 meters \citep{Jenniskens_2009,Farnocchia_2016,Farnocchia_2018}. Studying objects like these and understanding their physical properties will allow development of efficient mitigation strategies in the case of future life threatening impacts. In addition, observing small asteroids over long periods of time can allow for the characterization of size-dependent evolutionary processes. Specifically, the Yarkovsky and Yarkovsky-O'Keefe-Radzievskii-Paddack (YORP) effects \citep{Bottke_2006} can provide important information about asteroid spin, thermal, and/or interior properties \citep{Hanus_2018}. Lastly, small asteroids may have different physical properties than larger ones. Models for size sorting of surface particles via seismic shaking suggest that small bodies can have different surface particle size distributions than larger bodies \citep{Maurel_2016}. Efficiency of different regolith formation processes might be size dependent \citep{Delbo_2014}, and whether or not small objects are even covered by regolith is still debated. In general, asteroids larger than about 200 meters are not found to rotate faster than 2.2 hours \citep{Holsapple_2007}, though there are rare exceptions \citep{DeLuise_2007, Chang_2016, Polishook_2016}. Smaller asteroids however can rotate much faster, with some a rapid as 20 seconds per cycle \citep{Thirouin_2018}. These differences in spin properties  indicate that the internal structure of large and small objects could be different. While larger objects are usually considered to be rubble-piles, smaller ones could either be monolithic or possess sufficient internal strength to prevent them from breaking apart due to the centrifugal acceleration imparted by rapid rotation \citep{Herique_2018, Rozitis_2014,Polishook_2017a}.

In this work, we present visible spectra for 210 small NEOs (mean size around $D=60$m) observed in the framework of MANOS. The spectra of small NEOs allows us to derive their taxonomic classifications \citep{Bus_2002,Demeo_2009}. 
NEOs primarily originate from the Main Asteroid Belt \citep{Granvik_2018}, thus by understanding NEOs we probe the population of small MBAs which are currently inaccessible with current observational techniques. Comparing physical properties across size regimes both within and across populations may provide insight into size dependent evolutionary processes.

In the next section of this paper we present the observations of our 210 NEOs, the three facilities used for these observations, and our reduction procedure. Section \ref{sec.taxonomy} introduces asteroid taxonomy and describes the procedures used in this work. In Section \ref{sec:Result} we describe the properties of our sample in terms of absolute $H$ magnitude and equivalent estimated diameter, and discuss their taxonomic distribution. Section \ref{sec:biases} is devoted to the discussion of the different biases that might affect our sample. Section \ref{sec:Discussion} is devoted to the discussion of the results obtained by merging our sample with two other visible spectroscopic surveys of NEOs \citep{Binzel_2019,Perna_2018}. This allows us to analyze the largest available visible spectroscopic database covering asteroids from kilometer down to meter scales. We will discuss the properties of this sample in terms of the taxonomic distribution as a function of orbital parameters, MOID (Minimum Orbit Intersection Distance), and size.

\section{Observations and Data reduction}
\label{sec:obs}
The observations presented here were conducted over 5 years from August 2013 to March 2018 using both 8.1~m Gemini North (Mauna Kea, Hawaii, USA; MPC code:~568) and South (Cerro Pach\'{o}n, Chile; MPC code:~I11) \citep{Mountain_1994}, and Lowell Observatory's 4.3 m Discovery Channel Telescope (DCT; Happy Jack, Arizona, USA; MPC code: G37) \citep{Debring_2004}. The GMOS-N and GMOS-S (Gemini Multi Object Spectrometer) instruments \citep{Davies_1997} were used at the Gemini observatories and the DeVeny spectrograph \citep{Bida_2014} was used at the DCT. 

All observations were reduced using the same python-based spectral reduction pipeline optimized for asteroid spectral reduction. The pipeline was developed for this project and will be the focus of a future publication and public release. In the first step of the pipeline, each image is bias and flat field corrected. Biases are constructed by taking the median of a series of 5 to 11 zero-second exposures. The flat fields were acquired by uniformly illuminating a screen in the dome. A master flat field is constructed by first removing the spectral response of the lamp by normalizing each column (spatial direction) to the median. To avoid differential spatial variation with wavelength, the median is computed only around the region were the target spectra are located on the science images. Next, a cosmic ray filter is applied. We use the \textit{cosmic.py} python based cosmic detection and removal procedure\footnote{\url{https://obswww.unige.ch/\~tewes/cosmics\_dot\_py/cosmics.py\_0.4/doc/index.html}}. This code is based on the Laplacian cosmic ray detection algorithm by \citet{Van_2001}. For both GMOS instruments a spatial nodding procedure is employed during the observations. This technique involves taking spectral exposures with the target nodded to different spatial locations along the slit, and then subtracting pairs of exposures from one other to remove a majority of the sky background. However, due to changing sky conditions from one exposure to the next, some telluric emission lines remain after pair subtraction. A secondary step of background subtraction is then applied by fitting the residual background on either side of the target to interpolate the value at the location of the spectrum. In the case of the DeVeny spectra, no spatial nodding was used and only the second sky background subtraction method was applied. Each spectrum is then extracted, wavelength calibrated, and combined. The final step consists of dividing the NEO spectrum by the spectrum of a solar analog. The solar analog is observed immediately before or after the NEO and is chosen to match as closely as possible the NEO airmass. During the division step, the spectrum of the solar analogue is gradually shifted (shift of the order of $10^{-5}$ $\mu \rm{m}$) with respect to the spectrum of the asteroid in order to find the combination which provides the best correction of the telluric lines. Finally the spectrum is binned to a resolution of $\sim 200$ ($0.003$ $\mu \rm{m}$ bins). The pipeline also determines a spectral taxonomic classification by comparing the final asteroid spectrum with Bus-Demeo templates for each taxonomic class using a chi-square analysis. However, for consistency with previous surveys, the reported taxonomic classification was determine using the M4AST taxonomic classification webservice (\S\ref{sec.taxonomy} for more details). Table \ref{tab:obs} summarizes all the observations presented in this work.

\subsection{GMOS@Gemini}

We obtained 178 spectra of NEOs using the Gemini Multi-Object Spectrographs (GMOS) in the long-slit mode at both 8.1 m Gemini North (134 objects) and South (44 objects) telescopes. These instruments provide spectral observations from 0.36 to 0.94 $\rm{\mu}$m. 

In 2017, GMOS-North had a detector upgrade which provided better sensitivity in both the red and blue end of the spectral coverage. The old detector consisted of three 2048x4608 chips arranged in a row. Each of these detectors was an e2v deep depletion device with enhanced blue and red sensitivity. These detectors provided a plate scale of 0.0728 arcsec per pixels in the spatial direction and a dispersion of 0.174 nm per pixel for the R150 grating and 0.067 nm per pixel for the R400 grating. The upgraded array uses three 2048x4176 Hamamatsu detectors which are each optimized for throughput at their respective wavelength regimes.  The new plate scale is 0.0807 arcsec per pixel in the spatial direction with a dipsersion of 0.193 nm per pixel for the R150 and 0.074 for the R400 grating. The new Hamamatsu detectors were used for 6 targets observed at Gemini North in this work. In the case of Gemini South, all the spectra presented here were obtained with Hamamatsu detectors similar to those at Gemini North. The differences in resolution, detectors, and/or gratings across instruments had no siginificant affect on our final asteroid spectra, largely because we re-bin the final spectra by a factor of approximately 30 to decrease resolution and increase the signal-to-noise ratio of our faint targets. Such coarse binning effectively cancels the subtle differences across the instruments and detectors.

All spectra were acquired using the same observing sequence. Each target was observed with 6$\times$300 seconds individual exposures. Both GMOS instruments, either before or after upgrade, are multi-CCD detectors which cause small gaps in wavelength coverage. To obtain continuous wavelength coverage over the full 0.36-0.94 $\rm \mu$m range, the grating angle in the instrument was changed to produce a dispersion offset of 10 nm between the first three and last three exposures. For each grating offset, three spatial nods separated by 15 arcsec along the slit were used to enable sky background subtraction by taking the difference of pairs of images. Before or after each observation of an NEO target a solar analog standard star was observed using the same observation sequence to correct for the solar spectral component and telluric features. After the first 3 spectral exposures one flat field was acquired with identical grating angle and telescope pointing as used for the target. Then, a second flat field was acquired using the second grating angle before the final three spectral exposures of the target were obtained. Bias images and arc calibrations using a Ne-Ar lamp were acquired during the day before or after the observations. Two different gratings, 150 (R150) and 400 (R400) lines per mm, were used based on availability on the telescope for a given night. 

\subsection{DeVeny@DCT}

The third instrument we employed was the DeVeny spectrograph at Lowell Observatory's 4.3 m DCT. The DeVeny spectrograph was first known as the KPNO White Spectrograph at the Kitt Peak National Observatory (KPNO). It was acquired by Lowell Observatory in 1998 and used with the 72" Perkins telescope from 2005 to 2015, after which it was modified and installed on the DCT instrument cube \citep{Bida_2014}. The DeVeny spectrograph is equipped with a 2048x512 e2v CCD42-10 with 13.5 $\rm \mu$m pixels. It was operated using a grating of 150 lines per mm prodiving a dispersion of 0.43 nm/pixel and covering a spectral range from 0.32 to 1 $\mu \rm{m}$. The same reduction procedure was used for DeVeny data as for GMOS with only a few exceptions: no spectral or spatial nodding was performed when observing and no cosmic ray cleaning was needed during reductions. In total 32  NEOs were observed with this instrument.

\subsection{Data validation}

To validate our reduction pipeline and observation strategies, we compared our results with observations acquired with other instruments by other teams (Table~\ref{tab:Tax_Comp}). We found two objects that were also observed in the visible by the NEOSHIELD2 project \citep{Perna_2018} and six objects that were observed by the MITHNEOS project in the NIR \citep{Binzel_2019}. Table \ref{tab:Tax_Comp} summarizes taxonomic classifications in the visible, IR, and visible+near infrared (VISIR) spectral ranges for the two NEOSHIELD2 objects, the six MITHNEOS objects, and three other objects from the literature.

For the NEOSHIELD2 objects, we find the same taxonomic type for one (K-type; object 2015 XE) while the second, 2015 TM143, was found here to be Cgh versus Cb by NEOSHIELD2. This difference may simply be due to the low quality of our data at short wavelengths, which precludes detection of a spectral downturn short-ward of $0.5~\mu \rm{m}$ that can be taxonomically diagnostic. However, even though these are two different types, they correspond to the same complex. 

\begin{figure*}
\centering
\begin{tabular}{cc}
\includegraphics[width=8cm]{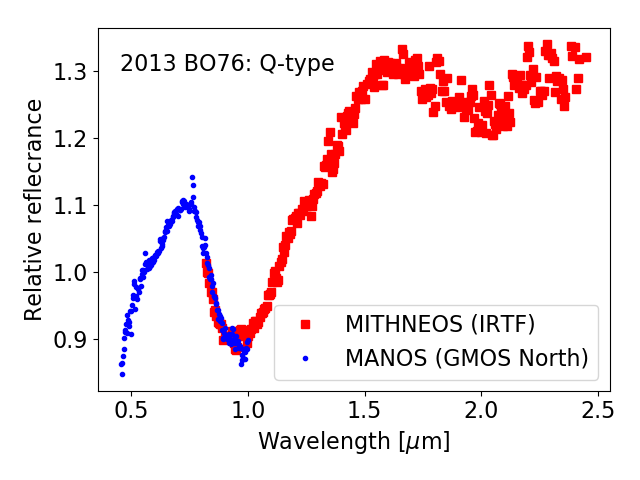}
     &
     \includegraphics[width=8cm]{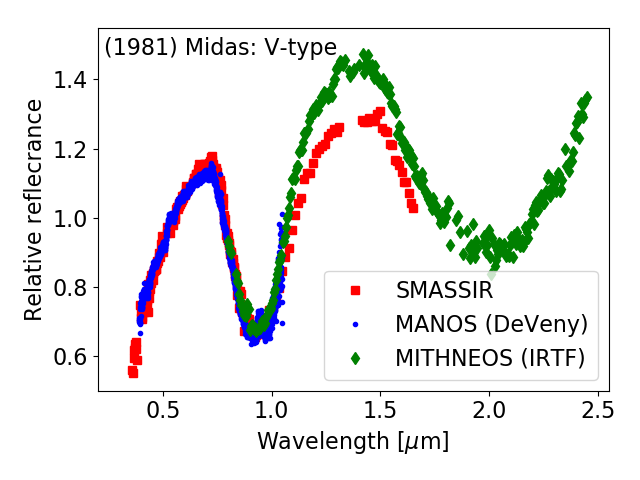}
\end{tabular}
\caption{MANOS Data validation. Left: comparison between the results from Gemini-GMOS (blue dots) and IRTF-Spex (red squares)  in the case of 2013 BO76. Right: comparison between the results from DCT-DeVeny (blue dots), SMASSIR (red squares), and IRTF-Spex (green diamonds) for the case of (1981)~Midas.  }
\label{fig:2013BO76_Deveny_MITHNEOS}
\end{figure*}

For the MITHNEOS data, even though these observations were not acquired in the same wavelengths regime as MANOS, we were able to compare our results by constructing a composite VISIR spectrum. In all cases the merging between the red end of the visible (GMOS) and the blue end of the NIR (MITHNEOS) spectra are in very good agreement. Figures \ref{fig:2013BO76_Deveny_MITHNEOS} shows comparisons between our observations, the MITHNEOS survey using the IRTF telescope, and the SMASSIR survey \citep{Burbine_2002}. We can see that all data generally agree with one other, though there are some slope differences in the NIR, possibly due to phase angle effects. 

Taxonomic classification of VISIR spectra used the MIT classification web service\footnote{\url{http://smass.mit.edu/cgi-bin/busdemeoclass-cgi}}.
For two out of the six MITHNEOS cases (2014 RC and 2014 SF304), we obtain identical classification. For two others (2013~PJ10 and 2013 BO76), \citet{Binzel_2019} reported several possible classifications and ours match with one of these. For the last two: 2010 CF19 is found to be within the same complex (X-complex) while 2014 WF201 is fit with two different complexes (Xc for MANOS and Ch for MITHNEOS). In general, these overlapping results across surveys are broadly consistent with one another within the limitations (signal-to-noise, wavelength coverage) of each data set. We note that the more comprehensive VISIR classification can differ from the visible- or IR-only classifications. Thus for the purposes of our analysis and to facilitate consistent comparisons across data sets (\S\ref{sec:Discussion}), we only consider from here onwards NEOs classified using visible data only and a single classification technique (\S\ref{sec.taxonomy}).

As a further validation step for the DeVeny spectrograph and our reduction pipeline, we observed a few well-studied objects (Table \ref{tab:Tax_Comp}). For (1036)~Ganymed and (1981)~Midas we obtained very good agreement with previously published taxonomic classifications (Sr and V respectively). In the case of (3752)~Camillo we found an Sr-type asteroid whereas \citet{deLeon_2010} found an Ld-type. However,  the NIR data obtained by these authors was not included in their taxonomic assignment and seems to indicate an S-complex object as opposed to an Ld-type. In addition polarimetric data (unpublished by M. Devogele) indicates an S-type classification. Asteroid 2008 EZ5 is the one object studied here with inconsistent classifications. The composite VISIR spectrum suggests an Sq classification while the individual spectra suggest different classes (Cg for the visible and X or D for the infrared). It is worth noting that the VISIR spectrum, even though classified as a Sq-type, does not match well with the Sq reference, and the near-IR component is relatively low signal-to-noise. 

\begin{deluxetable*}{c|cccccc}
\tablecaption{Comparison of the taxonomic classification for objects observed by MANOS and other surveys. References: 1 \citet{Binzel_2019}, 2 \citet{Binzel_2004}, 3 \citet{Perna_2018}, 4 \citet{Binzel_2001}, 5 \citet{deLeon_2010}   \label{tab:Tax_Comp}}
\tablehead{
\colhead{Object} &Instrument& \colhead{VIS (This work)} & \colhead{VISIR (This work)} & \colhead{VIS} & \colhead{IR}  &  \colhead{VISIR} }
\startdata
(1036)~Ganymed 	& DeVeny &Sr	&  Sr		&  S$^2$		& Sr 			& Sr$^1$ \\ 
(1981)~Midas	     & DeVeny	&V		& V		&	V$^4$		&		V			&	V$^1$	\\
(3752)~Camillo		& DeVeny &Sr    &    &           &          Ld$^5$          &    \\ 
2008 EZ5 				& DeVeny &Cg	&  Sq		&  	& X/D$^1$				&\\
2010 CF19 			& GMOS-N &Xc	& Xc		&  	& C/X$^1$				&\\
2013 BO76 			& GMOS-N &Q	&  Q		& 				&S/Sq/Q$^1$ &\\
2013 PJ10 			& GMOS-N &Sr	&  Sq		&  	& S/Sr$^1$  	&\\
2014 RC 				& GMOS-S &Sq	& Sq		& 	& Sq$^1$				&\\
2014 SF304		 	& GMOS-N & Q 	& 	Q	&   		& Q$^1$				&\\
2014 WF201 			& GMOS-N & Xc	&  Xe		& 	& Ch$^1$			&\\
2015 TM143 			& GMOS-N &Cgh 	&  & 	Cb$^3$		& 					&\\
2015 XE 				& GMOS-N &K 	&  	& K$^3$			& 					&\\
\enddata
\end{deluxetable*}

\section{Taxonomic classification} \label{sec.taxonomy}

Taxonomic classification is used to group asteroids based on the characteristics of their spectra. There are several taxonomic classification schemes which have been developed using different data sets covering different spectral ranges and resolutions. Taxonomic classification roughly differentiates between common mineralogical classes present in the asteroid population. Here we make use of the Bus-DeMeo taxonomic classification system \citep{Demeo_2009}, primarily developed for visible plus near-infrared wavelengths (0.45 to 2.45 $\rm \mu$m). Even though our dataset does not cover near-infrared wavelengths, the Bus-DeMeo taxonomy is amongst the most comprehensive and is very similar to the visible-only Bus and Binzel system \citep{Bus_2002}. Moreover, our spectra go beyond the 0.82 $\rm \mu$m limit defining the Bus and Binzel system. In order to make use of the extra wavelength coverage (up to 1 $\rm \mu$m for DeVeny), the use of the Bus-DeMeo taxonomic classification system is needed. 

For the remainder of this work we will primarily consider just taxonomic complexes as opposed to individual types. This allows better statistics (e.g. more objects per group), moreover, the distinction between types inside a complex is based on subtle spectral variations (slope, shallow absorption bands) that can only be properly resolved in high signal-to-noise spectra, which is not always the case here. We define the S complex as the collection of spectra belonging to the S, Sr, Sv, and Sk-classes. We do not include the Sq class in the S-complex, as done by \citet{Demeo_2009}, but rather in a Q-complex combining the Q and Sq-classes as defined by \citet{Binzel_2004}. The reasons for this are the very high fraction of Q and Sq-types amongst NEOs compared to the MBA population (objects on which the \citet{Demeo_2014} system was based) and the correlation of Q and Sq-types with low degrees of space weathering (\S\ref{sec:S_Q}). In addition, we do not include the L, Ld, and K-classes in the S-complex as was done by \citet{Binzel_2004}, because these types are likely compositionally distinct from the S-complex \citep{Devogele_2018, Sunshine_2008}. The Ld class does not exist in the Bus-Demeo taxonomy. 
We combine the K and L-classes into the K complex as these two classes are barely distinguishable at visible wavelengths. NIR data are needed to clearly discriminate these two classes. 
Finally, we define the C-complex as the group of the B, C, Cb, Cg, Ch, and Cgh-classes, and the X-complex as the group of the X, Xc, Xk, and Xe-classes. In each of these five complexes (S, Q, K, C, X) we have 35, 57, 18, 23, and 50 objects respectively in the MANOS sample. On the other hand, the A, D, O, R, T, and V are defined as individual classes and are not included to any larger complex.

In the following sections, we will compare our results with those obtain by \citet{Perna_2018} as part of the NEOSHIELD2 project and by \citet{Binzel_2004, Binzel_2019} as part of the MITHNEOS survey. The NEOSHIELD2 database consists of 146 visible spectra classified in the Bus-DeMeo system as determined by the webservice M4AST\footnote{\url{http://spectre.imcce.fr/m4ast/index.php/index/start}}. This early version of the MITHNEOS database contains visible spectra only, and principal component analysis \citep[PCA, e.g.][]{Bus_2002} was used to classify their spectra. We are aware that more recent databases of NEO spectra exist \citep[e.g.][]{Binzel_2019}, however the majority of these newer data are near-infrared spectra only and thus do not compare directly to the MANOS sample. The \citet{Binzel_2004} sample also includes non-NEOs that are Mars crossers, which we exclude from our analysis. In total we considered 286 spectra from the \citet{Binzel_2004} dataset. 

In order to compare the results from these different sources, one further step is needed. Comparing taxonomic classifications obtained from different techniques can lead to erroneous statistics. Ours and the NEOSHIELD2 dataset have been analysed using a chi-square technique. This involves finding a best fit to template spectra of each class by minimizing a chi-square statistic. On the other hand, the  \citet{Binzel_2004,Binzel_2019} spectra have been classified using a PCA method. To compare the three databases, we re-determined the taxonomy of all the spectra presented in \citet{Binzel_2004,Binzel_2019} using the M4AST webservice. Surprisingly we find that using the chi-square fitting method on the \citet{Binzel_2004,Binzel_2019} dataset leads to significant variations in the fraction of S-complex (as defined by \citet{Demeo_2009}) asteroids: from $52.9 \pm 4.3$\% for PCA to $36.0 \pm 3.5$\% for chi-squared. On the other hand, the fraction of Q-type asteroids increased from $6.2 \pm 1.5$\% for PCA to $15.6 \pm 2.3$\% for chi-squared, A-type from $0.35\pm 0.35$\% to $1.7 \pm 0.8$\%, O-type from $1.7 \pm 0.7$ to $3.1 \pm 1.0$\%, K-type from $2.4 \pm 0.9$\% to $5.2 \pm 1.3$\%, and L-type from $2.4 \pm 0.9$\%  to $3.8 \pm 1.1$\%. The uncertainties on the reported fraction were computed by taking into account Poisson statistics on the number of spectra for each class. If we sum the increases in Q, A, O, K, and L-types we retrieve the fraction of S-complex asteroids previously determined by the PCA method. This suggests that the chi-squared technique distributes objects with 1 $\mu \rm{m}$ absorption features across a greater diversity of spectral types than PCA. As such chi-squared-derived taxonomic classifications will show an increase in the number of Q, A, O, K, and/or L types (at the expense of S-types) relative to PCA. We also see a decrease of the X-complex fraction from $15.9 \pm 2.3$\% to $10.8 \pm 1.9$\% and an increase of the C-complex from $7.3 \pm 1.6$\% to $12.1 \pm 2.0$\% when comparing chi-square to PCA. These findings clearly demonstrate the need for an homogeneous taxonomic classification scheme. Thus, we also determine the taxonomic classification of our data-set using the M4AST webservice (which uses a chi-square method analogous to our pipeline). As expected, we find that the taxonomic classifications provided by our pipeline are very similar to those obtained with M4AST. Considering Poisson statistic, the difference between the number of objects in the different class or complex stays bellow 0.5 $\sigma$ for half of them.

\section{Results} \label{sec:Result}

\begin{figure*}[t!]
\centering
\includegraphics[width=12cm]{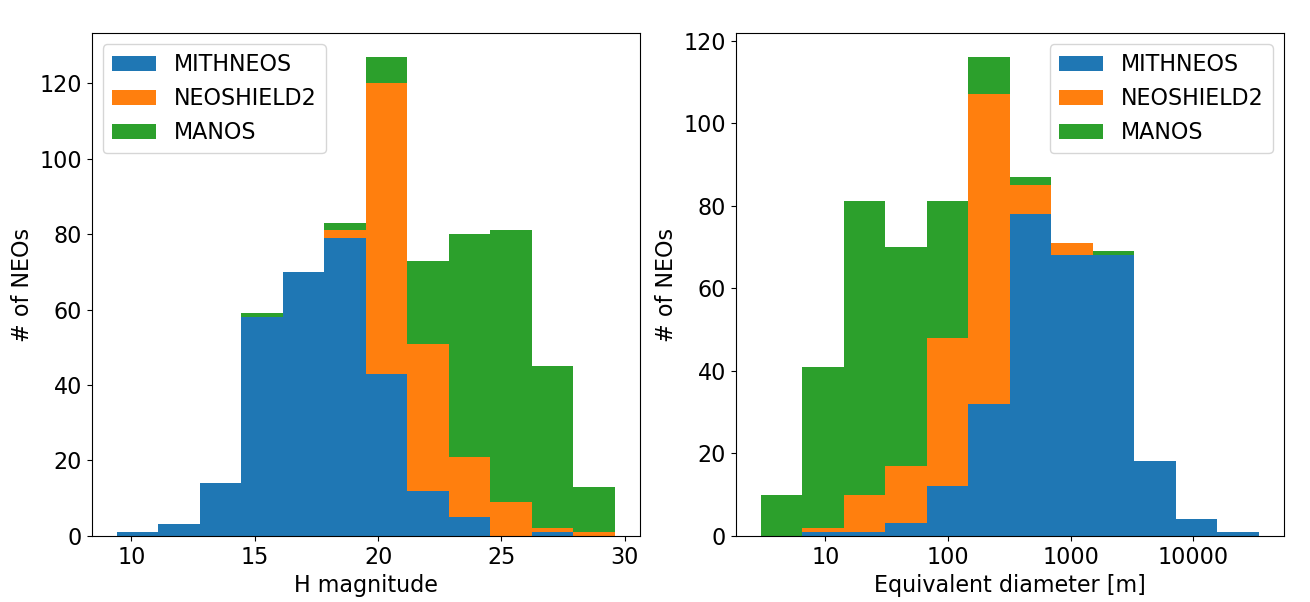}
\caption{Stacked histograms of the H magnitude (left) and equivalent diameter (right) distributions in the MANOS, NEOSHIELD2, and MITHNEOS databases. MANOS generally probes the smallest asteroids, MITHNEOS the largest, and NEOSHIELD2 is intermediate.}
\label{fig:H_Histo_All}
\end{figure*}

We report here the taxonomic classification of 210 NEOs observed in the framework of the MANOS project. This database contains approximately 3\% of the currently known population of NEOs with size $D<100$ m. The distributions of $H$ and equivalent diameter $D$ of the objects in the MANOS database are displayed in Figure~\ref{fig:H_Histo_All}. The $H$ magnitude has been converted into equivalent diameter $D$ considering the average albedo ($p_{\rm V}$) for each taxonomic class as reported by \citet{Thomas_2011a}. The mean $H$ value of our data set is around 25th magnitude. 
Taking into account the expected albedo for each taxonomic type and the $H$ magnitude of each object, the mean equivalent diameter of all objects in the MANOS dataset is around $D$ = 50 meters, with the smallest objects expected to be as small as 3 meters.    

\begin{figure}[ht!]
\plotone{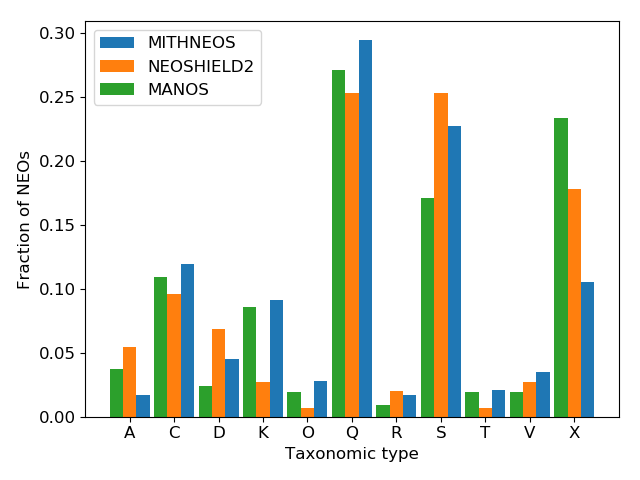}
\caption{Taxonomic distribution of NEOs in the MANOS, NEOSHIELD2, and chi-squared reclassified MITHNEOS spectroscopic databases in relative percentage. The fraction of S-complex NEOs is lower in the MANOS database while the fraction of X-complex is higher in MANOS and NEOSHIELD2 compared to MITHNEOS. \label{fig:Tax_Bar_Frac}}
\end{figure}

The taxonomic distribution in our dataset is reported in Table \ref{tab:Tax_Dist} and as bar plots in Figure~\ref{fig:Tax_Bar_Frac} in relative fraction for each individual taxonomic class or complex. The classes which are the most represented are the Q, X, S, C, and K complexes with respectively $27.1 \pm 3.6$ \%, $23.8 \pm 3.4$ \%, $16.7 \pm 2.8$ \%, $10.9 \pm 2.3$ \%, and $8.6 \pm 2.0$ \% of the full population.

In Figure \ref{fig:Tax_Bar_Frac} we compare our results with those of NEOSHIELD2 \citep{Perna_2018} and our chi-squared re-classification of the \citep{Binzel_2004,Binzel_2019} spectra. The NEOSHIELD2 dataset contains 146 objects with a mean $H$ magnitude of 22 and mean equivalent diameter of  180 m (3 times larger than the MANOS sample). The mean $H$ magnitude of the Binzel dataset is 17.7 mag corresponding to  an equivalent diameter of approximately 1 kilometer (20 times larger than the MANOS sample). These three datasets are highly complementary and sample very different size regimes in the NEO population with MANOS providing the majority of spectra for $H>23$ or $D<100$ m. These samples also differ in their orbital element distributions (\S~\ref{sec:biases}). The full combined MANOS+MITHNEOS+NEOSHIELD2 sample contains 642 spectra in roughly equal proportion across the three surveys.

In Table~\ref{tab:Tax_Dist}, we report the number of spectra and relative fraction in each taxonomic class or complex for each dataset. We also report the deviation between our data and the NEOSHIELD2 and MITHNEOS datasets. This deviation corresponds to the fractional difference between the two surveys divided by the uncertainty on this difference computed by uncertainty propagation. The main difference we find is a net decrease in the fraction of S complex NEOs in the MANOS dataset. The fraction of S complex asteroids is  $22.7 \pm 2.8\%$ in the MITHNEOS and $26.0 \pm 4.2\%$ in the NEOSHIELD2 datasets while it is only $16.7 \pm 2.8\%$ in MANOS. These numbers correspond to differences compared to MANOS of 1.7 and 1.8 sigma for the NEOSHIELD2 and MITHNEOS datasets, respectively. Hypotheses to explain these differences will be discussed in \S~\ref{sec:S_Q}.  In addition, we see a net increase in the fraction of X-complex asteroids with $10.8 \pm 1.9\%$, $17.2 \pm  3.4\%$, and $23.8 \pm 3.4\%$ in the MITHNEOS, NEOSHIELD2, and MANOS datasets respectively. This corresponds to an increase of 2.9 sigma between MITHNEOS and MANOS. X-complex asteroids will be discussed in detail in \S~\ref{sec:A_X}. The fraction of K-complex in the MITHNEOS survey is comparable with the fraction observed by MANOS while NEOSHIELD2 observed a much smaller fraction. However, due to the very shallow absorption band characterizing the K-complex at visible wavelengths, this result could be related to differences in wavelength coverage betweem instruments and surveys. 

\begin{deluxetable*}{l|cr|crr|crr}
\tablecaption{Taxonomic distribution of objects presented in this work compared with those of the NEOSHIELD2 and MITHNEOS databases  \label{tab:Tax_Dist}}
\tablehead{
\colhead{Taxonomy} & \colhead{\#} & \colhead{Fraction} & \colhead{\#} & \colhead{Fraction} & \colhead{Dev.}  & \colhead{\#} & \colhead{Fraction} & \colhead{Dev.} \\
\colhead{} & \colhead{} & \colhead{\%}  & & \colhead{\%} & \colhead{$\sigma$}  & & \colhead{\%} & \colhead{$\sigma$}}
\startdata
	&  &	MANOS ~   		 &		 & NEOSHIELD2	& 		~	& ~	&		MITHNEOS					&	~	\\ \hline
A & 8 & 3.8 $\pm$ 1.3 & 8 & 5.5 $\pm$ 1.9  & +0.7 &  5 &  1.7  $\pm$ 0.8 &  -1.3 \\
C (C, Cg, Cgh, Ch, Cb, B) & 23 & 10.9 $\pm$ 2.3 & 14 & 9.6  $\pm$ 2.6 & -0.4 & 34  & 11.9 $\pm$ 2.0 &  +0.3\\
D & 4 & 1.9 $\pm$ 0.9 & 10 & 6.8  $\pm$ 2.1 & +2.1 & 13 &  4.5 $\pm$ 1.2 & +1.7 \\
K (K, L)& 19 & 9.0  $\pm$ 2.1 & 5 & 3.4 $\pm$ 1.5 & -2.2 & 25 & 8.7 $\pm$ 1.7   & -0.1  \\
O & 4 & 1.9 $\pm$ 0.9 & 1 & 0.7 $\pm$ 0.7 & 1.0 & 8 & 2.8 $\pm$ 1.0 & +0.7 \\
Q (Q, Sq)& 57 & 27.1 $\pm$ 3.6 & 37 & 25.3 $\pm$ 4.2 & -0.3 & 84 & 29.4  $\pm$ 3.2& +0.5 \\
R & 2 & 0.9 $\pm$ 0.7  & 3 & 2.0 $\pm$ 1.2 & +0.8 & 5 & 1.7 $\pm$  0.8 &  +0.8  \\
S (S, Sa, Sr, Sv)& 35 & 16.7 $\pm $ 2.8  & 38 & 26.0 $\pm$ 4.2 & +1.8 & 65 & 22.7 $\pm$ 2.8 & +1.5   \\
T & 4 & 1.9 $\pm$ 0.9  & 1 & 0.7 $\pm$  0.7 & -1.0 & 6 & 2.1 $\pm$ 0.9 &  +0.1  \\
V & 4 &  1.9 $\pm$ 0.9 & 4 & 2.7 $\pm$ 1.4  & +0.5 & 10 &  3.5 $\pm$ 1.1 & +1.1 \\
X (X, Xc, Xe, Xk, Xn) & 50 &  23.8 $\pm$ 3.4 & 25 & 17.2  $\pm$ 3.4 & -1.4 & 31 & 10.8 $\pm$ 1.9 & -3.3  \\
\enddata
\end{deluxetable*}

\section{Observational biases \label{sec:biases}}

Biases are inherent to any survey, either intentional as with the MANOS focus on small size and low $\Delta v$, or unintentional like the discovery bias toward high albedo objects (see \citet{Granvik_2018} for a detailed discussion about the discovery bias of NEO discovery surveys). $\Delta v$ in this context is defined as the impulse needed for a spacecraft to maneuver from low Earth orbit to a rendezvous with the asteroid in its orbit. It can be computed for NEOs using the approximation described by \citet{Shoemaker_1978}. In this section we discuss several of these biases to understand their effect on the observed population and taxonomic distribution of NEOs measured by each of the surveys discussed in this work. A more detailed de-biasing of our sample will be the focus of a future publication.

\subsection{Bias towards high albedo} 
The first bias is a discovery and observational bias towards high albedo asteroids. When observed at visible wavelengths, for similar sizes, high albedo asteroids are brighter and can be more easily discovered and characterized. This bias leads to an observational preference for high albedo classes such as O, A, Q, or S, and under-observation of low albedo classes such as D or C. Discovery bias by the Catalina Sky Survey, currently the predominant NEO discovery survey in the world, has been extensively discussed in \citet{Granvik_2018}.

According to \citet{Stuart_2004}, the de-biased fraction of S-complex NEOs is 22\% while MITHNEOS observed 31.8 $\pm$ 3.2\% (S+K complexes reported here) and MANOS 25.7 $\pm$ 3.5\%. We also note that the de-biased fraction of Q complex is estimated to be around 14\% while MITHNEOS observed 29.4 $\pm$ 3.2\% and MANOS 27.1 $\pm$ 3.6\% . The S and Q complexes are represented by high albedo asteroids (respectively 0.26 and 0.29 according to \citet{Thomas_2011a}). On the other hand low albedo classes such as D types ($p_{\rm V} = 0.02$) have an de-biased population estimated to be around 17\% while only 4.5 $\pm$ 1.2\% and 1.9 $\pm$ 0.9\% were observed by MITHNEOS and MANOS respectively. These statistics clearly show the expected over-observation of high albedo and under-observation of low albedo asteroids. 

\subsection{Bias toward low MOID} 

\begin{figure}[ht!]
\plotone{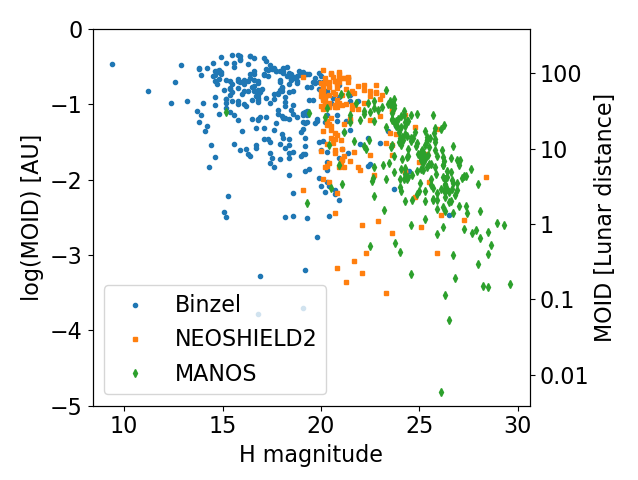}
\caption{Plot of the absolute $H$ magnitude as a function of the MOID for asteroids in the MITHNEOS, NEOSHIELD2 and MANOS databases \label{fig:H_MOID}. High MOID asteroids are no longer observed as the $H$ magnitude increases}
\end{figure}

Bias toward high albedo is not the only bias present in our sample. MANOS focuses on small objects, which due to their intrinsic faintness are necessarily low MOID objects in order to be observable. This introduces a strong selection effect that biases our observed taxonomic distribution relative to other surveys. The mean MOID for the MITHNEOS, NEOSHIELD2, and MANOS surveys are respectively 0.113, 0.083, and 0.016 AU. Figure \ref{fig:H_MOID} is a simple illustration of this MOID bias. It shows a plot of the absolute $H$ magnitude as a function of the logarithm of the MOID in AU for all asteroids considered in this work. We can clearly see that as $H$ magnitude increases, high MOIDs are no longer observed. 

Low MOIDs may have consequences for the surface properties of these asteroids. These asteroids, by definition, are making close approaches to the Earth. These approaches can result in tidal forces that induce surface rejuvenation, effectively suppressing the effects of space-weathering \citep{Binzel_2010}. This would then suggest an increase in the fraction of fresh, unweathered spectral types in the MANOS data set. In \S\ref{sec:S_Q} we discuss possible reasons for why this is not the case.

\subsection{Orbital elements biases} 

\begin{figure*}[ht!]
\plotone{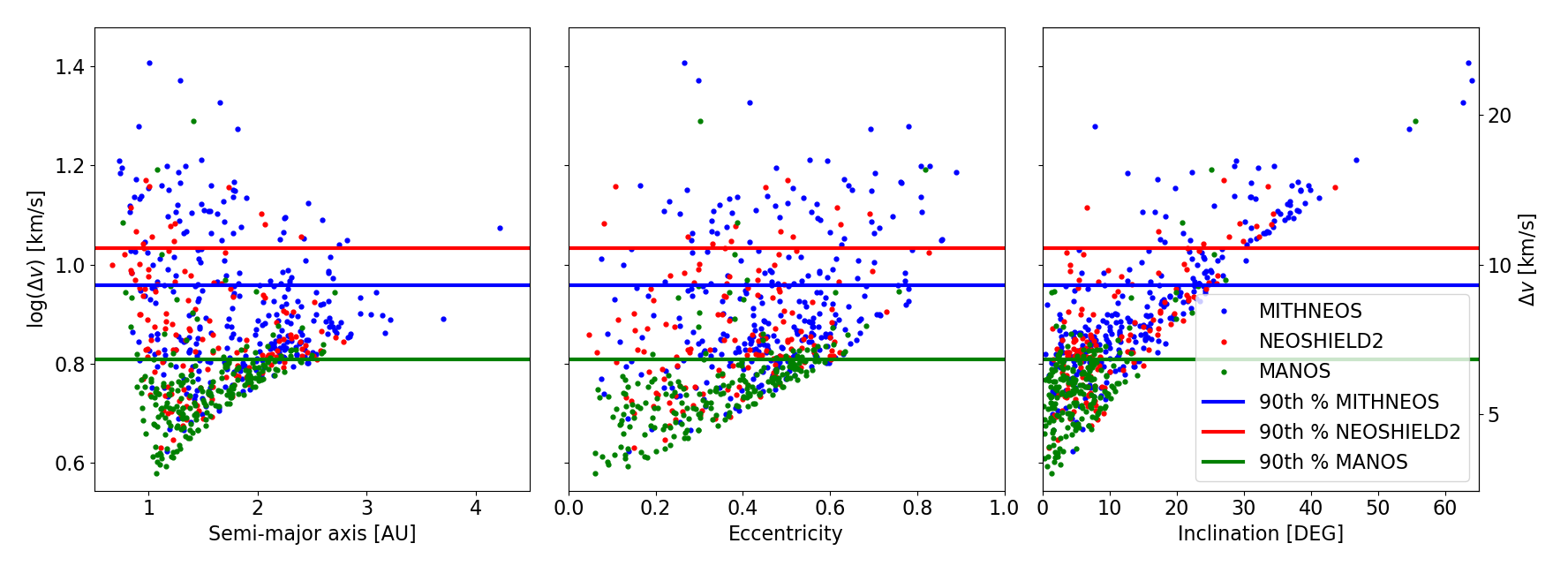}
\caption{From left to right: plots of the semi-major axis $a$, eccentricity $e$, and inclination $i$ as a function of $\Delta v$. The different colors represent the different surveys: MITHNEOS (blue), NEOSHIELD2 (red), and MANOS (green). The lines represent the 90th percentile in $\Delta v$ for each survey (respectively 9.1, 10.8, and 6.4 km/s). We note that MANOS observed objects in a narrower range of orbital element space ($0.85<a<2.6$ AU, $e<0.62$, and $i<16^{\circ}$) than the other surveys.\label{fig:DELTAV_AEI}}
\end{figure*}

The MANOS survey focuses on low MOID, low $\Delta v$ asteroids. Observing exclusively low $\Delta v$ asteroids is introducing a bias towards semi-major axis around $a=1$ AU, low eccentricity $e$, and low inclination $i$. This bias can be seen in Figure \ref{fig:DELTAV_AEI} which shows the semi-major axis $a$, eccentricity $e$, and inclination $i$ as a function of the $\Delta v$ for each survey considered in our analysis. This plot also shows the 90th percentile $\Delta v$ limits for the different surveys  which correspond to $\Delta v = 9.08$, 10.78, and 6.44 km/s for the MITHNEOS, NEOSHIELD2, and MANOS surveys, respectively. We can see that by focusing on low $\Delta v$ objects, MANOS ignores high eccentricity and high inclination targets, while the semi-major axes of our targets are generally closer to that of the Earth compared to other surveys. 

\begin{figure}[ht!]
\plotone{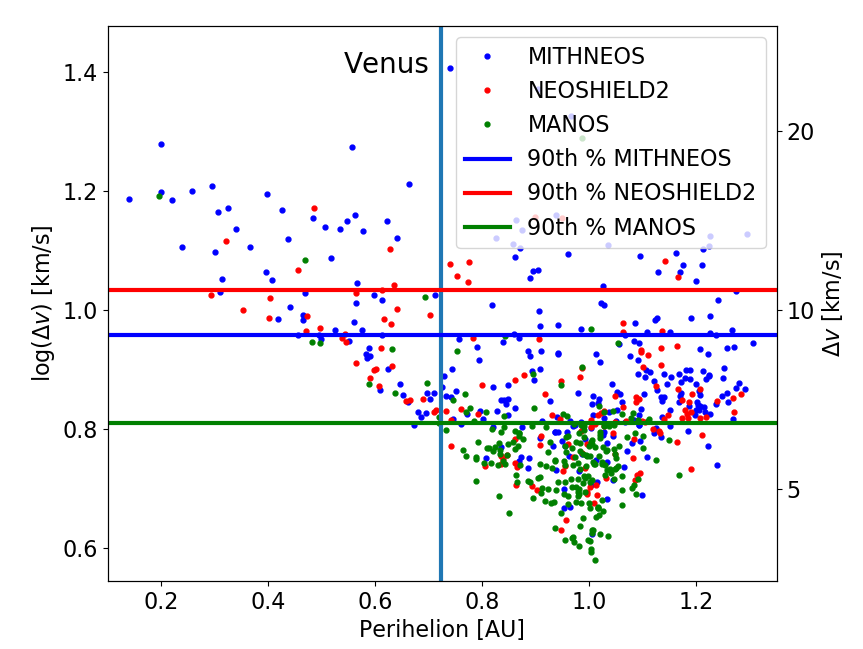}
\caption{Perihelion as a function of the $\Delta v$ for each survey (MITHNEOS: blue, NEOSHIELD2: red, and MANOS: green). The horizontal lines represent the 90th percentile of $\Delta v$ for each survey. The vertical line corresponds to the semi-major axis of Venus a=0.723 AU. MANOS only observed targets with perihelia $0.73 < q < 1.17$ AU. \label{fig:Perihelion_DeltaV}}
\end{figure}

The MANOS bias toward Earth-like semi-major axis and low eccentricity naturally introduces a bias toward Earth-like perihelia. Figure \ref{fig:Perihelion_DeltaV} represents the perihelion distance of all the objects considered in this work as a function of their $\Delta v$. As with Figure \ref{fig:DELTAV_AEI}, the horizontal lines correspond to the 90th percentile $\Delta v$ lines for the individual surveys. The vertical line corresponds to the semi-major axis of Venus. Note that the 90th percentile $\Delta v$ of MANOS coincidentally corresponds to the semi-major axis of Venus. This means that MANOS targets have a lower probability of making a close encounter with Venus than objects observed in the other surveys. For example, assuming $a=1$ AU and $i =0^{\circ}$, the median $\Delta v$ for each survey corresponds to minimum perihelia of 0.57, 0.67, and 0.78 AU for MITHNEOS, NEOSHIELD2, and MANOS respectively (lower perihelion could be reachable for the same $\Delta v$ considering lower semi-major axis values, however objects with $a<1$ only represent 9\% of all the objects considered in this work). This bias towards Earth-like perihelia also introduces a bias towards higher Venus MOID in the MANOS sample. The fraction of objects with Venus MOID smaller than 0.02 AU is respectively 5.2, 5.5, and 0.9\% in the MITHNEOS, NEOSHIELD2, and MANOS surveys. Similar to the process that occurs due to Earth encounters \citet{Binzel_2010}, a low MOID to Venus is expected to increase the likelihood of planetary encounters and thus the chance for surface re-freshening events that can affect spectral type. The effect of having a low MOID to Earth and Venus is important while considering the S and Q-complexes. This will be discussed in detail in Sec.~\ref{sec:S_Q}. 

These orbital element biases also introduce a bias toward specific source regions in the Main Belt. Near Earth Asteroids originate from different regions in the Main Belt known to be taxonomically heterogeneous \citep{Demeo_2014}. NEO orbital elements contain a vestige of their original Main Belt source region, thus their escape region probability can be determined \citep{Granvik_2018}. \citet{Granvik_2018} considered six different escape regions for the Near-Earth Asteroids -- the Hungaria and Phocaea clusters, the $\nu_6$ and the Jupiter resonances 5:2, 3:1, and 2:1 -- and provided the orbital steady state distribution of the NEOs originating from these regions. The MANOS survey is biased towards Earth-like perihelia and low inclination. The Phocaea region is characterized by a relatively small semi-major axis and eccentricity, but has very high inclination ($i \sim 30^{\circ}$). Since only 5\% of the MANOS objects have inclination higher than $14^{\circ}$, objects from the Phocaea region should be very rare in the MANOS sample. On the other hand, 47\% and 27\% of the objects in the MITHNEOS and NEOSHIELD2 surveys respectively have an inclination higher than $14^{\circ}$. In the case of the MANOS sample, the same bias against Phocaeas is also true for the Hungaria region, but to a lesser extent since their inclination is smaller ($i \sim 20^{\circ}$). 
The $\nu_6$, 5:2, 3:1, and 2:1 source regions originate at semi-major axis around 2.2, 2.5, 2.7, and 3.2 AU respectively. Thus, at face value, the most likely source region for MANOS targets seems to be the $\nu_6$ resonance: our dataset includes only 6.5\% of objects with $a>2.2$ AU, 2\% with $a>2.5$ AU, and only one object with $a>2.7$ AU. However, the low MOID values characteristic of our sample suggest that MANOS objects may have an increased likelihood to have lost the memory of their source regions due to close interactions with Earth. As such it may be non-trivial to unambiguously determine the source regions for some MANOS objects. A detailed analysis of this issue is beyond the scope of this work.

\subsection{Bias due to asteroid size}

The main difference between the three surveys discussed in this work is the size of the observed objects. The median $H$ magnitude is respectively 17.5, 21, and 25 mag for the MITHNEOS, NEOSHIELD2, and MANOS samples. The fraction of objects coming from different Main Belt source regions is dependent on size \citep{Granvik_2018}. Table \ref{tab:SR} summarizes the fraction of NEOs coming from each source region according to \citet{Granvik_2018} for $H$ lower than 17.5, 21, and 25 mag. Even though these statistics are representative of the cumulative distribution, the number of objects in the population increase so quickly with $H$ that for a given $H$ cutoff larger objects become negligible in terms of their contribution to the presented fractions. This can be seen for the Phocaea objects which decrease from 7 to 0.02\% for $H<17.5$ and $H<25$ mag. 

Table \ref{tab:SR} indicates that objects from the Phocaea cluster, and the Jupiter resonances of 2:1, and 5:2 are negligible contributors to the MANOS sample, whereas they account for 29 and 17\% respectively in the MITHNEOS and NEOSHIELD2 surveys. The main increase in the MANOS sample comes from the Hungaria region which increases by a factor of 2.1 compared to the MITHNEOS sample, and 4.8 compared to the NEOSHIELD2 survey. The 3:1 fraction increases by a factor of 1.9 compared to the MITHNEOS survey and is similar to that for NEOSHIELD2. The $\nu_6$ fraction is similar for MANOS and MITHNEOS, and increases by 10\% for the NEOSHIELD2 sample. The taxonomy distribution in each of these source regions is different and since the contribution of each one is directly dependant on size, the taxonomy distribution of the NEO population should also be dependant on object size. These variations of the taxonomy distribution with size will be discussed in detail in \S\ref{sec:S_Q} for the Q and S complexes, and \S\ref{sec:A_X} for the A-type and X-complex.

\begin{deluxetable}{cccc}
\tablecaption{Fraction, in \%, of NEOs coming from the different source regions according to \citet{Granvik_2018} for targets with $H$ smaller than 17.5, 21, and 25 mag corresponding to the median $H$ for the MITHNEOS, NEOSHIELD2, and MANOS survey respectively. \label{tab:SR}}
\tablehead{
\colhead{Source region} &\colhead{$H<17.5$} & \colhead{$H<21$} & \colhead{$H<25$}}
\startdata
$\nu_6$ & 39 & 49 & 38  \\
3:1J & 19 & 31  & 36  \\
5:2J & 17 & 12  & 0.1 \\
Hungaria & 11 & 5 & 24   \\
Phocaea & 7 & 3  & 0.02   \\
2:1J & 5 & 2 & 0.1 \\
\enddata
\end{deluxetable}

\section{Discussion} \label{sec:Discussion}

In this section the observed variation of the taxonomic distribution in the MANOS database compared to the NEOSHIELD2 and MITHNEOS will be discussed. While combining these three different data-sets, trends are also observed as a function of size (H magnitude), MOID, or orbital elements. 

The first two sections will focus on specific classes or complexes while the last section will focus on one specific mechanism. Section \ref{sec:S_Q} is devoted to the S and Q complexes. The total fraction of S+Q complexes is observed to decrease as a function of size. Mechanisms allowing S to turn onto Q-complex asteroids will be discussed. The Q/S ratio is found to vary as a function the Earth and Venus MOID as well as perihelion distance. Section \ref{sec:A_X} will discuss A-type and X-complex for which a relative increase compare to other taxonomic type/complex is observed as a function of size. The last section is about the size/density dependent disaggregation of asteroids \citep{Scheeres_2018}. This only mechanism can possibly explain the overall observed variation of the taxonomy distribution as a function of size.

\subsection{S and Q-complex asteroids \label{sec:S_Q}}

S- and Q-complex asteroids are compositionally related. Q-type asteroids have been linked to the fresh surface of ordinary chondrite meteorites \citep{Mcfadden_1985, Nakamura_2011}. The surface of such an asteroid, when exposed to the space environment, sees its reflectance properties change due to space weathering. For ordinary chondrites, the effects of space-weathering include an increase in spectral slope, a lowering of albedo, and a reduction of absorption band depth. These processes turn Q-types to S-type asteroids \citep{Chapman_1996}.

\begin{figure}[ht!]
\plotone{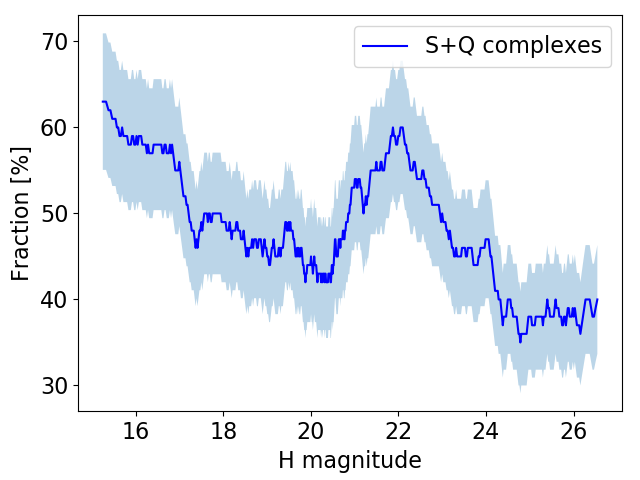}
\caption{Running mean of the number of S+Q complex asteroids across the MANOS, NEOSHIELD2, and MITHNEOS samples as a function of $H$ magnitude. The shaded area corresponds to the uncertainties taking into account Poisson statistics. This change could be due to size dependent differences in Main Belt source region and/or size dependent surface properties like grain size. \label{fig:QS_Sum}}
\end{figure}

The fraction of S-complex asteroids is significantly lower in the MANOS sample compared to MITHNEOS and NEOSHIELD2. We examine all objects with ordinary chondrite-like compositions by combining the fractions of the S- and Q-complexes which represents 52.1, 51.4, and 43.8\% respectively for the MITHNEOS, NEOSHIELD2 and MANOS samples. Figure \ref{fig:QS_Sum} represents the running mean of the observed S+Q  fraction as a function of H magnitude across all three surveys. It can be seen that the S+Q fraction goes from 60\% for $H=16$ mag down to 48\% for $H = 17.4$ mag, and 42\% for $H =25$ mag. We also see an interesting peak around $H$=22 mag with a fraction of 60\%. The decrease of the S+Q fraction as a function of $H$ can tentatively be explained by a variation of the source regions of the objects as a function of size. Since the Phocaea asteroids are primarily composed of S-complex asteroids \citep{Carvano_2001} and their abundance amongst NEOs rapidly decreases with size (Table \ref{tab:SR}), we expect to find of order 7\% less S+Q-complex asteroids for $H<25$ mag compared to the fraction for $H<17.5$ mag (assuming Phocaea's are 100\% S-complex). This corresponds well to the 8.3\% decrease in S+Q asteroids observed in the MANOS sample compared to MITHNEOS. The higher fraction of S around $H=22$ mag could be explained by the higher fraction of asteroids coming from the $\nu_6$ (10\% more than for $H<17.5$ mag and 11\% more than for the MANOS sample). The implication of this interpretation for the non-uniform S+Q fraction is that source regions in the Main Belt can produce taxonomic or compositional variation within the NEO population that is size dependent. Our analysis suggests that this may be an observable signature.

Another tentative explanation for the S+Q fraction trend could be that as body size decreases, surface properties, such as the mean grain size, change. All taxonomic classification systems have been defined based on spectra of MBAs or large NEOs, which are expect to have surfaces dominated by small grain sizes \citep{Jaumann_2012,Robinson_2001} that likely dominate the optical properties at visible and near-infrared wavelengths. Such taxonomic systems may break down when considering significantly different grain size regimes, for example in the nearly regolith free surface of NEO Ryugu. Evidence for changes in surface grain size as a function of object diameter has been seen with such in-situ observations of NEOs \citep{Dombard_2010, Tancredi_2015, Michikami_2018}. Indirect evidence for the possibility of different surface properties includes the significantly different centripetal accelerations on the surfaces of small objects, where rotation periods less than 20 seconds have been observed \citep{Thirouin_2018}. Such rapid rotation could have implications for the retention of small grains on the surfaces of these bodies. Note that the peak around $H=22$ mag in Figure \ref{fig:QS_Sum} closely corresponds to the transition from purely gravity dominated rubble piles to bodies where cohesion can play a significant role in dictating internal structure and strength \citep{Scheeres_2010}. However, it is not clear why a peak in the S+Q fraction would occur at this transition. 

Size-dependent changes in surface properties were also predicted by models for objects with low planetary MOID (i.e. those experiencing frequent planetary encounters like in the MANOS sample). These objects can experience gravitationally induced seismic shaking, which can affect surface grain size distributions \citep{Maurel_2016}. The implication of these interpretations of the S+Q fraction is that NEOs can have fundamentally different surfaces as a function of size. This can be directly tested with additional in situ spacecraft observations of NEOs across a range of sizes, and telescopic observations that can constrain surface grain properties such as measurements of thermal inertia \citep[e.g.][]{MacLennan_2018,Hanus_2015} or polarimetric phase curves \citep[e.g.][]{Cellino_2018, Devogele_2018}.

In order to analyze the role of planetary encounters in surface alteration, we consider the Q/S ratio. One model for S-type asteroids to turn into Q-type involves surface re-freshening during planetary encounters \citep{Nesvorny_2005}. In order to make a close planetary encounter, an object must have a low MOID relative to the planet. This model is supported by our data since the Q/S ratio for the MANOS survey is Q/S = 1.6 $\pm$ 0.3, while it is only 1.3 $\pm$ 0.2 and 1 $\pm$ 0.2 for the MITHNEOS and NEOSHIELD2 surveys respectively. This is expected because MANOS is biased toward low MOID asteroids. However, these Q/S ratios are only marginally significant in their difference when taking into account uncertainties based on the size of each sample. One can also compares the fractions of Sq and Q sub-type objects (here Q-types is considered as its individual class without taking into account the Sq objects). We see that the Q/Sq ratios for the MANOS, NEOSHIELDS2, and MITHNEOS surveys are equals within the uncertainties with respectively 0.8 $\pm$ 0.2, 0.8 $\pm 0.3$, and 1.1 $\pm$ 0.3.

\begin{figure*}[ht!]
\plotone{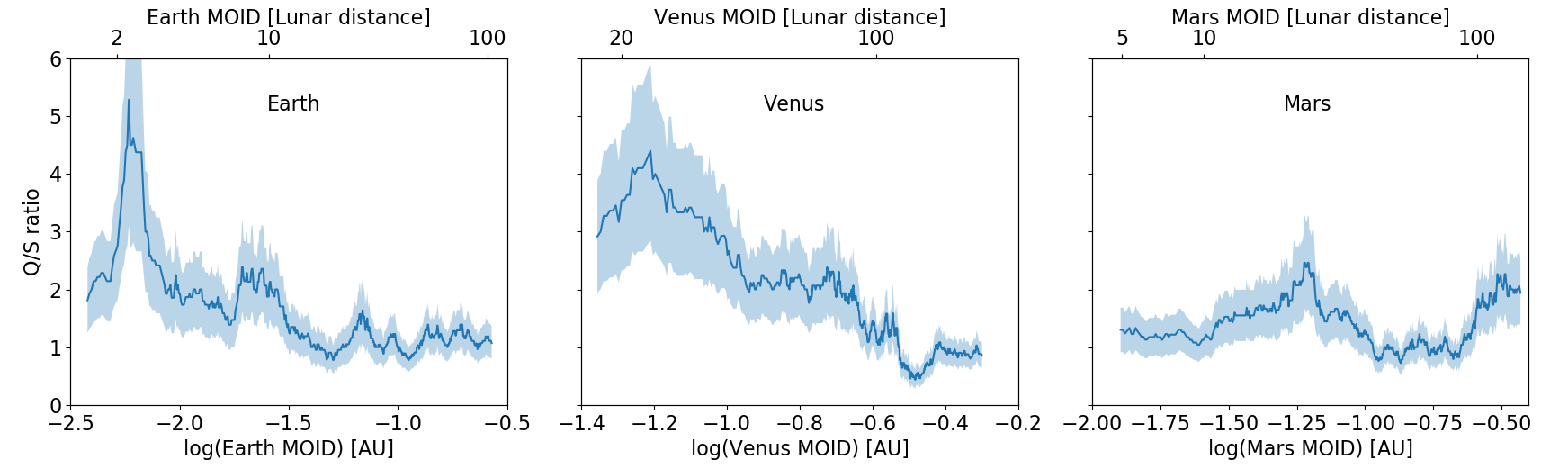}
\caption{Running mean of the ratio between the number of Q and S asteroids across the MANOS, NEOSHIELD2, and MITHNEOS samples as a function of the Earth, Venus, and Mars MOID (respectively from left to right). The shaded areas correspond to the uncertainties considering Poisson statistic for the S and Q complexes fractions and uncertainties propagation. \label{fig:MOID_Other_Q_S}}
\end{figure*}

Figure \ref{fig:MOID_Other_Q_S} represents the running mean of MOID, over the full sample, in bin sizes of 100 asteroids as a function of the Q/S ratio for the Earth, Venus, and Mars. In the case of Earth, the Q/S ratio does not go to zero for the highest MOID, but stabilizes around a 1:1 ratio at MOID $>0.1$ AU. These high MOID asteroids are too distant to experience close encounters with Earth. Including Mars-crosser asteroids in this analysis, the Q/S ratio decreases to 0.33, but does not go to zero. By definition these objects have a MOID larger than 0.3 AU and have no interaction with Earth. Either interactions with Mars or some other surface refreshing mechanism that doesn't involve planetary encounters (e.g. collisional origin in the Main Belt, or YORP spin-up) are likely responsible for the Q-types in this Mars crosser population \citep{Demeo_2014}. We note that while a very close planetary approach ($<1$ lunar distance) is actually needed to refresh the surface of an asteroid, this plot only considers instantaneous MOID. As pointed-out by \citet{Binzel_2010}, Q-type asteroids with MOID as large as 0.15 AU can have a MOID as small as $10^{-5}$ AU in the past $10^{5}$ years. They also noted that not all asteroids with very low MOID in the recent past are Q-types. This fact is relevant to the following discussion on Venus MOID, Mars MOID, and perihelion distance. Considering Q and Sq-type as separate classes, we observe that the Q/S and Sq/S ratio as a function of MOID distances, for each planets, are well withing the error bar associated to each curves. Thus, no differences can be seen between Q and Sq-types. In other words, the Sq/Q ratio remains constant for any MOID within error bars.



Asteroids with high Earth MOID can also have low MOID relative to other planets such as Venus. The effectiveness of a planetary body to refresh an asteroid surface depends on the mass of the planet. This makes Venus nearly as effective as the Earth, while Mars is expected to be much less efficient. The middle panel of Figure \ref{fig:MOID_Other_Q_S} represents the running mean of the Q/S ratio as a function of Venus MOID. We can see that as for the Earth the Q/S fraction increases with smaller MOID. We can also see that the increase starts further from the planet than for Earth. This is interpreted as the fact that an asteroid, observable from the Earth, which would have a low Venus MOID can also have a low MOID relative to the Earth, increasing the probability of planetary encounters. On the other hand, the right hand panel of Figure \ref{fig:MOID_Other_Q_S} represents the case of the Mars MOID where no increase of the Q/S ratio is seen. This is consistent with Mars being much less effective than the Earth and Venus in converting S to Q-type asteroids.

\begin{figure}[ht!]
\plotone{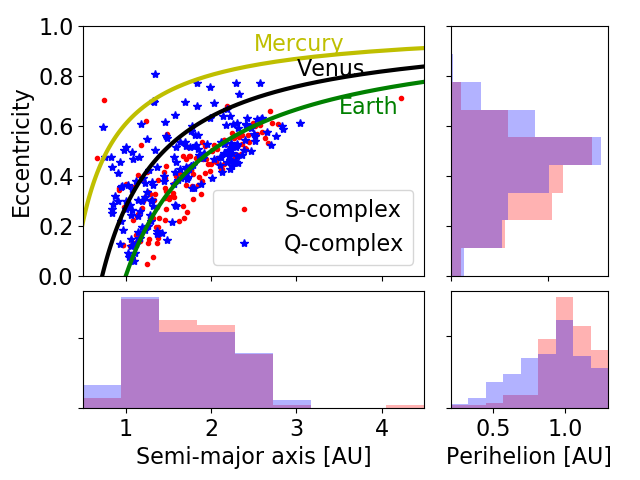}
\caption{Top left panel: Semi-major axis versus eccentricity plot for S (red dots) and Q-complex (blue stars) asteroids. Lines corresponding to perihelion equal to the semi-major axis of the Earth (1 AU), Venus (0.723 AU), and Mercury (0.387 AU) are displayed in green, black and yellow respectively. Other panels: distribution in eccentricity (top right), semi-major axis (bottom left), and perihelion (bottom right) of the Q (blue) and S (red) complexes. The locations where both histograms overlap are displayed in purple. An excess of Q-complex objects is most pronounced at small perihelion distances. \label{fig:Q_S_A_E_Peri}}
\end{figure}

Related to these MOID relationships, the Q/S ratio is also dependant on orbital elements. The top left panel of Figure~\ref{fig:Q_S_A_E_Peri} represents the semi-major axis $a$ versus eccentricity $e$ for S and Q-complex asteroids. While the S-complex asteroids predominantly remain near the 1 AU perihelion line, a non-negligible fraction of the Q-complex asteroids plot well above this line. The other panels of Figure~\ref{fig:Q_S_A_E_Peri} represent the distribution in semi-major axis (bottom left), eccentricity (top right), and perihelion distance (bottom right) for the S and Q complexes. A Kolmogorov-Smirnov (KS) analysis rejects with 99.9\% confidence the null hypothesis that the eccentricity and perihelion distributions of Q and S complex asteroids are drawn from the same distribution (Table \ref{tab:KS}). On the other hand, the same test for the semi-major axis shows that the null hypothesis can only be rejected with 63\% confidence. This means that the distribution of semi-major axis for Q and S is the same, but they are significantly different in terms of eccentricity and hence perihelion as well.
Another important observation of Figure~\ref{fig:Q_S_A_E_Peri} is that the relative fraction of S-complex asteroids is rapidly decreasing with lower perihelion distance while the relative fraction of Q-complex is not. This rapid decrease of S-complex objects coincides with perihelion distances inside of Venus' orbit. This suggests the intriguing possibility that Venus, in addition to the Earth, may play a role in the generation of Q-type asteroids.

The same KS test of comparing the perihelion distribution of the S-complex (N=138) to those of the Sq (N=91) and Q sub-types (N=87) (treated as separate classes) shows that the null hypotheses can be rejected, for both, with a confidence higher than 99\%. However comparing the perihelion distribution of Sq with that of the Q sub-type shows that the null distribution can only be rejected with 89\% confidence, suggesting that these two distributions are more likely to come from the same population. We also find that the confidence level for rejection is higher when comparing S-complex perihelia to  Q-types than when comparing the S-complex to Sq-types. Comparing the eccentricity distributions, we find that the distributions of the S-complex and Sq-type are likely to come from the same population while the populations of the S-complex and Q-type are different. Overall these results are suggestive of a continuous transition from the S-complex to the Sq-type and then to the Q-type as a function of eccentricity and (by extensive) perihelion. In the case of semi-major axis, all distributions are identical.
\begin{deluxetable}{c|cccc}
\tablecaption{Two-sided KS probability for combinations of perihelion and eccentricity distributions from the S-complex (S, Sa, Sr, Sv), Q-complex (Sq, Q), Sq-types (Sq), and Q-types (Q). The numbers represent the probability of rejection of the null hypothesis in \%. \label{tab:KS}}
\tablehead{& \colhead{S-complex} &\colhead{Q-complex} & \colhead{Sq-type} & \colhead{Q-type} \\
& \colhead{N= 138} &\colhead{N=178} & \colhead{N=91} & \colhead{N=87}} 
\startdata
& & Perihelion & & \\
\hline
S-complex & 0 & 99.998 & 99.76 & 99.99 \\
Q-complex & 99.998 & 0  & - &  -\\
Sq-type & 99.76 & -  & 0  & 89 \\
\hline
& & Eccentricity & & \\
\hline
S-complex & 0 & 99 & 58 &  99.9 \\
Q-complex & 99 & 0  & - &  -\\
Sq-type & 58 & -  & 0  & 94 \\
\enddata
 \end{deluxetable}

\begin{figure}[ht!]
\plotone{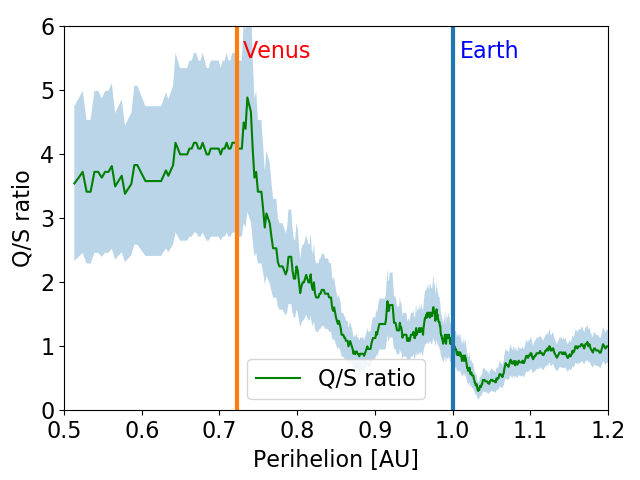}
\caption{Running mean of the Q/S ratio in 100 object bins as a function of perihelion distance. The shaded areas correspond to the uncertainties considering Poisson statistic for the S and Q complexes fractions and uncertainties propagation. The semi-major axis of the Earth (1 AU) and Venus (0.723 AU) are shown by blue and red lines respectively. We can see that the Q/S ratio changes slope at perihelion distances around Earth and Venus. We interpret the large increase between Earth and Venus as a mutual influence of both planets modifying surfaces through planetary encounters. \label{fig:Q_S_Peri}}
\end{figure}

As such we consider the running mean of the Q/S ratio as a function of perihelion distance (Figure \ref{fig:Q_S_Peri}). As expected the Q/S ratio increases with lower perihelion. However, this increase is not linear, but has two distinct changes in slope corresponding closely to the perihelion distances of Earth and Venus. We also note that there is a plateau for objects with perihelion between 0.88 and 1 AU. These objects encounter Earth, but stay far away from Venus. The maximum Q/S ratio is around 4.9 for perihelia near the semi-major axis of Venus. This suggests that Venus plays a role in the refreshening of S-complex asteroid surfaces. The minimum Q/S ratio is 0.30 at a perihelion distance of 1.03 AU, just outside of the Earth's orbit. The cause of the modest increase in this ratio at perihelion distances greater than 1.03 AU is unclear. Mars does not appear to have the same effect that Earth and Venus do in resurfacing objects during close encounters (Figure \ref{fig:MOID_Other_Q_S}). It is possible that objects recently escaped from the Main Belt are Q-types, for example following collisional removal from a precursor parent body, and thus are affecting the Q/S ratio at larger perihelion distances. 

We saw previously that the MANOS survey's focus on low $\Delta v$ objects causes a bias against low Venus MOID objects compared to the NEOSHIELD2 and MITHNEOS surveys (Figure \ref{fig:Perihelion_DeltaV}). As Venus seems to play a role towards increasing the Q/S ratio, this bias should produce a smaller Q/S ratio in the MANOS sample, whereas the MANOS bias towards lower Earth MOID targets should act in the opposite manner. The relative importance of these biases in the MANOS sample remains unclear.

The non-zero Q/S ratio for asteroids which do not interact with Earth or Venus could be explained in several ways:

\begin{itemize}
    \item MOID is not a static value but instead evolves with time. Asteroids with a large MOID to Venus and Earth today might have had much lower MOIDs in the past. Backwards orbital integrations, similar to those performed by \citet{Binzel_2010}, could lend insight into this possibility, but is beyond the scope of this work. 
    \item The fresh Q-type surfaces could come from a collision with another asteroid. However, the collision probability in near-Earth space is much lower than in the Main Belt. Asteroids would then have to enter NEO space as Q-types instead of S-type. A counter would be that very few Q-complex objects have been found in the MBA, though the studied size regimes are very different. Spectra for MBAs are mainly for objects with $D>5$ km while most of the NEOs have $D<5$ km. It was also found that small objects from recent dynamical families are more likely to display a fresh Q-type surface \citep{Thomas_2011b}.
    \item Asteroids can experience spin rate changes due to the YORP effect. This acceleration can cause surface material to migrate towards the equator or even escape the surface \citep{Walsh_2008}, thus exposing fresh un-weathered terrain. Indeed, some asteroids pairs that were recently formed by the rotational-fission process were found to display fresh Q-type surfaces \citep{Polishook_2014}. Combining spectral and rotational data could provide insight into this possibility.
    \item Regolith formation processes such as thermal fatigue \citep{Delbo_2014} are expected to be independent of MOID and are also strongly dependent on perihelion distance. This could help to explain the background of Q-complex objects in near-Earth space. 
\end{itemize}

\subsection{X-complex and A-type asteroids \label{sec:A_X}}

The fraction of X-complex and A-type asteroids increases significantly in the MANOS and NEOSHIELD2 surveys compared to the MITHNEOS survey. 

\begin{figure}[ht!]
\plotone{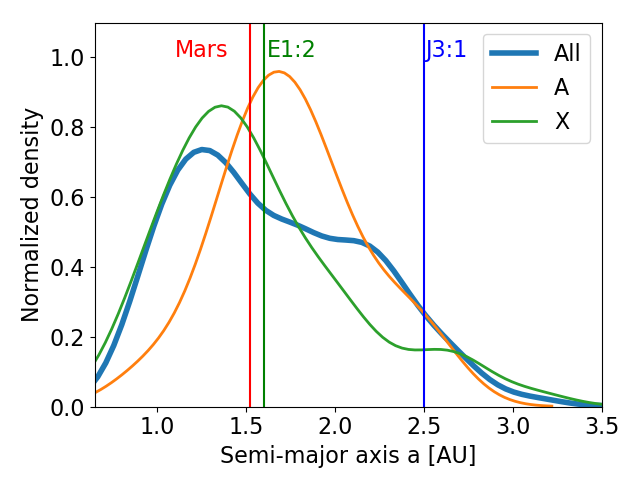}
\caption{Density distribution of NEOs for the A-types and X-complex as a function of semi-major axis. The red vertical line corresponds to the semi-major axis of the orbit of Mars. The green and blue vertical lines correspond to the location of the 2:1 Earth and 3:1 Jupiter mean motion resonances.  \label{fig:Density_a}}
\end{figure}

Figure \ref{fig:Density_a} shows the density distribution of objects in the A-class and X-complex as a function of semi-major axis $a$. The blue curve represents the full dataset and serves as a reference. This curve is characterized by a bi-modal distribution with a main peak at 1.275 AU and a secondary peak at 2.1 AU, corresponding to the inner edge of the Main Belt. There are only two escape regions with $a<2.1$ AU for objects to leave the Main Belt and enter the NEO population. These correspond to the Hungaria and Phocaea asteroid families. Objects in these families can be destabilized by mean motion resonances with Mars and Jupiter, and to a lesser extent Earth, Venus, and Saturn \citep{Mceachern_2010}. Several secular resonances such as the $\nu_5$, $\nu_6$, or $\nu_{17}$ can also play an important role in NEO delivery from these regions \citep{Warner_2009}. No significant resonances exist around 1.275 AU, whereas the peak at 2.1 AU corresponds to a peak of asteroids in the de-biased NEO population \citep{Granvik_2018}.

A-type asteroids are uncommon all over the main-asteroid belt. \citet{Demeo_2019} reported fractions of A-types for the inner, mid, and outer main belt to be 0.22\%, 0.14\%, and 0.11\% respectively. These values are more than 10 times lower than those found in the MANOS and NEOSHIELD2 database. However, the fraction of chi-squared classified A-types in the MITHNEOS sample is 4.9 times higher than the fraction of PCA classified A-types. Taking this into account, the PCA classified fraction of A-types in the MANOS sample is estimated to be $0.8 \pm 0.3$. This number is more in accordance, however still 4 times higher, with the estimated fraction of A-type in the Main Belt.

This over-abundance of A-types observed in the MANOS database could be an effect of the observation bias toward high albedo objects. A-types are high albedo ($\sim$0.3-0.5) objects \citep{Demeo_2019}. This effect is expected to be stronger for smaller size object which are harder to discover and characterize.

On the other hand, one has to keep mind that sub-kilometer asteroids cannot be observed in the Main Belt and their fraction cannot be observationally estimated. An indirect estimation by \citet{Turrini_2016} suggested that a fraction of 0.7\% sub-kilometer A-types in the Main Belt could explain olivine patches detected on the surface of Vesta.

If A-types are distributed more or less evenly throughout the main-belt, some regions in the Solar System show over-densities. In that matter, a cluster of 11 asteroids within the fifth Lagrangian point of Mars, was dynamically associated with the asteroid (5261)~Eureka, the largest object in the group \citep{Christou_2013}. Three of them were found to have A-type spectra (in both the visible and near-IR regime \citep{Borisov_2016,Polishook_2017b}). Based on the rarity of A-type asteroids, it is likely that these asteroids share a common heritage. \citet{Polishook_2017b} suggested this common origin involved impact ejection from Mars. The Hungaria region is also known to display an over-density of A-type asteroids. A recent spectroscopic survey by \citet{Lucas_2018} reported a relative A-type fraction of 1.5\% in the Hungaria region. An other estimates by \citet{Demeo_2013} suggested that 7\% of the mass of the Hungaria group is of A-type. However more recent studies \citep{Demeo_2019} suggests a much smaller value of 0.26\%.

The A-type population observed in the MANOS survey shows an excess of objects located at semi-major axes around 1.6 AU and is associated with a deficiency of objects below 1.3 and over 2.1 AU (see Fig. \ref{fig:Density_a}). The location of this peak corresponds with the 1:2 Earth mean motion resonance which provides the means for Hungaria family asteroids to enter the NEO region.

In addition, the fraction of A-types in the NEO region was found to be size dependent, increasing with smaller sizes \citep{Perna_2018}. This may be consistent with an increase of the fraction of A-type with smaller size in the main belt \citep{Turrini_2016} or the NEO source region models. As discussed earlier, the relative fraction of Hungaria asteroids in the NEO population increases from 11\% for $H<17.5$ mag to 24\% for $H<25$ mag. This is consistent with an increase of A-types by a factor of 2.3 between the MITHNEOS and MANOS surveys. On the other hand, the predicted fraction of Hungaria NEOs (A-types and others) is only 5\% for $H<21$ mag, which seems inconsistent with the high fraction ($\sim5\%$) of A-types in the NEOSHIELD2 sample.

Considering these observed properties of A-type asteroids, we suggest that the over-abundance of A-types in NEO space compared to the observed fraction in the Main Belt could be due to either an observation bias toward high albedo asteroids, an increase of the fraction of A-type for unobserved sub-kilometer A-type in the main-belt or a variation of the feeding source region of NEO as a function of size. In the latter scenario A-type NEOs would then be relic pieces of Mars, which is consistent with the fact that Martian meteorites have been found on Earth. The last hypothesis would imply the fact that some asteroids are mistakenly defined as A-type if they are only observed in the visible. \citet{Demeo_2019} found that only half of the A-types defined by the visible regime are actually A-types, when you observe them in the near-IR.

The X-complex, as with the A-types, also shows a strong peak at small $a$ (Figure \ref{fig:Density_a}). The Hungaria region is composed of roughly 56\% X-type asteroids \citep{Lucas_2018}. The X-complex regroups three old taxonomic classes which are distinguished by albedo $p_V$, but are otherwise spectroscopically similar. P-types are characterized by $p_{\rm V} < 0.14$, M-types have 0.14 $<p_V<$ 0.30, and E-types have albedo higher than 0.30 \citep{Tholen_1984}. X-types in the Hungaria region are almost exclusively E-type asteroids \citep{Lucas_2018}. 

From the 106 X-complex asteroids reported in this work, thirteen of them have an albedo measured by NEOWISE. Of these 13, two are consistent with an E-type classification, thus suggesting that $16 \pm 11$ of the X-complex objects in this study are E-type, representing $2.5 \pm 1.7$\%  of the full sample considered here. 
From the NEO population with $17<H<22$ mag (the X-types for which we have albedo information are exclusively in this range), 5.6\% come from the Hungaria region (Table \ref{tab:SR}). As only 56\% of Hungarias are E-type, the expected fraction of E-type asteroids in our sample is 3.1\%. This number is consistent with the $2.5 \pm 1.7\%$ suggested by the observations. Specific to MANOS, if the fraction of Hungaria asteroids increases to 24\% in the range $17<H<25$ mag, as suggested by \citet{Granvik_2018}, then we might expect to find many more E-types in the MANOS sample. Unfortunately only four MANOS targets have an albedo determination. Of those four, one is an X-type with a high albedo, i.e. is an E-type. A larger sample of small NEO albedos, particularly in the range $22<H<25$ mag, would serve to further test this predicted contribution of Hungaria E-types to the NEO population.

We also note that, only the X-types show a significant increase in the MANOS survey: their abundance increases by 61\% relative to MITHNEOS. However, the X-complex combines asteroids with a high diversity of albedos, from very dark P-types ($p_{\rm V} < 0.14$) to very bright E-types ($p_{\rm V} > 0.30$). The reported mean albedo of X-types by \citet{Thomas_2011a} is $p_{\rm V} = 0.31$, whereas \citet{Stuart_2004} reported only $p_{\rm V} = 0.06$. The bias toward observation of high albedo combined with the increase of high albedo X-type asteroids from the Hungarias region might explain the rapid increasing of X-type while observing smaller asteroids.  

\subsection{Catastrophic disaggregation of asteroids}

Several processes leading to full or partial disaggregation of asteroids have been suggested. First, \citet{Scheeres_2018} proposed a mechanism leading to the full disruption of small asteroids when reaching a spin rate threshold. The size limit at which such full dissagregation would occur is dependent on density ($R_0 \propto 1/\rho$). On the other hand, the time needed for a body to experience a full disaggregation is also proportional on the density. A low density asteroid will start to disaggregate at larger radius, but would take longer time than a high density one. We also note that $R_0 \sim 100$ m which corresponds to the transition between NEOSHIED2 and MANOS data. The effect on the observed population of object smaller than 100 m is still unclear, but it might result in variation of the fraction of asteroids of different compositions as a function of size and density.

Taking into account density estimations for the different taxonomic types \citep{Carry_2012}, we see that C and S-complex are the lowest density ($\sim 2.1$) while A-type and X and K-complex have relatively higher densities ($\sim 3.7$). For complexes, we report here the mean value of all type belonging to the complex (see \citet{Carry_2012} for individual estimation for individual types). These densities have been derived based on large main-belt asteroids which might not reflect the true densities of NEOs. However, even if the NEOs possess larger macro-porosity due to their smaller size, we are interested here in the relative variation of the density as a function of the taxonomic type and not the absolute values.

If MANOS is observing objects around this transition, then this effect could be a reason for the observed diminution of S-types relative to the higher density X, A, and K-types. Moreover, the size of an object component parts (i.e. boulders) might also be dependant on density, thus also contributing to different taxonomic distributions at small sizes.

The second mechanism involve the catastrophic disruption of asteroids due to solar heating at small perihelion distance \citep{Granvik_2016}. This mechanism is also size dependent, but also albedo, and perihelion distance dependent. 

The catastrophic disruption of an asteroid due to solar heating is expected to happen at perihelion distances too low to affect the MANOS dataset which mainly focus on asteroid with perihelion distance close to 1 AU. Even if the perihelion distance at which this catastrophic disruption occurs is dependent on the asteroid size, the average disruption distance pass from 0.06 AU to 0.18 for $H$ magnitude from 18 to 24, only 8 targets in MANOS dataset possess a $q < 0.7$ AU. Moreover, the perihelion distribution for high and low albedo asteroids for asteroids with $q>0.6$ AU is found to be similar while it is not while considering the full NEO population \citep{Granvik_2016}. Even if this analysis was made on large asteroids, this effect is expected to be negligible for Earth like perihelion MANOS target.

\section{Conclusions}
The MANOS project is a survey of small, low $\Delta v$, low MOID NEOs employing spectroscopic, photometric, and astrometric techniques. As part of this survey we present 210 new visible spectra. The mean $H$ of the MANOS database is around 25 mag which corresponds to a mean equivalent diameter of 50 m. In this paper we presented the taxonomic distribution of these objects. 

We compared the taxonomic distribution of the MANOS dataset with other datasets of visible NEO spectra (MITHNEOS and NEOSHIELD2). These two surveys sample asteroids that are generally larger than those studied by MANOS. Comparing the taxonomic distribution across surveys we find:
\begin{itemize}

\item The fraction of S-complex asteroids in the MANOS database of $16.7 \pm 2.8\%$ is lower than the fraction in the NEOSHIELD2 and MITHNEOS database which are respectively $26.0 \pm 4.2$ and $22.7 \pm 2.8\%$.  The reason for this decrease is unclear but could be due to an observational selection of lower MOID asteroids in the MANOS sample. The decrease of S-complex could be partially explained by a refreshing of the asteroid's surfaces during close approaches with Earth, Venus, and/or Mars, causing a change in surface features from S to Q-type. This is supported by the fact that the Q/S ratio is increasing in the MANOS database (1.3 in MITHNEOS, 1.0 in NEOSHIELD2, and 1.6 in MANOS).  
\item The fraction of S+Q-complex asteroids in the MANOS database of $43.8 \pm 4.6$\% is lower than the fraction in the NEOSHIELD2 and MITHNEOS database which are respectively $51.4 \pm 5.9$ and $52.1 \pm 4.2$\%. When combining these datasets this decrease goes from 60\% at $H=16$ mag to 42\% at $H=26$ mag. This could be explained by a decrease in the fraction of asteroids coming from the Phocaea region at small sizes. Another possible explanation would be changes in surface properties (e.g. grain size distribution) as a function of size.
\item The fraction of X-complex asteroids in the MANOS database of $23.8 \pm 3.4$\% is significantly higher than the fraction in the MITHNEOS database ($10.8 \pm 1.9$\%) and is about the same as in the NEOSHIELD2 database ($17.2 \pm 3.4$\%). This increase of X-complex asteroids is interpreted as both an increase in the fraction of NEOs from the Hungaria family at small sizes ($H>22$ mag) and a discovery bias towards high albedo E-type asteroids. The high fraction of Hungaria asteroids is supported by the fact that 2 out of 13 X-types with measured albedos are high, consistent with an E-type classification and a Hungaria origin. This seems to support NEO source region models \citep[e.g.][]{Granvik_2018} that predict NEOs of different sizes preferentially come from different parts of the Main Belt. However, this conclusion is based on low number statistics. An albedo survey of X-type NEOs would provide further insight into this scenario. 
\item The fraction of A-type asteroids in the MANOS database of $3.5 \pm 1.2$\% lies between the fraction reported by NEOSHIELD2 ($5.5 \pm 1.9$ \%) and MITHNEOS ($1.7 \pm 0.8$ \%). This increase compared to the MITHNEOS database is interpreted, like the X-types, as a combination of an increase of NEOs coming from the Hungaria region at small sizes and a bias toward high albedo objects. These A-types would then potentially be pieces of Mars ejected during the early stages of Solar System evolution. This could explain the higher fraction of A-types in the NEO population as compared to that found in the Main Belt.
\item We presented evidence that Ven us encounters play a role in the process of turning S-complex into Q-complex asteroids. We also notice that the Q/S ratio is highly correlated with asteroid perihelion distance. This correlation could be due to other physical processes like thermal fatigue which should increase with lower perihelion. However, discontinuities in this ratio are seen around perihelia equal to Earth and Venus, suggesting multiple processes may be at play.
\end{itemize}   

\section{Acknowledgements}

These results made use of the Discovery Channel Telescope at Lowell Observatory. Lowell is a private, non-profit institution dedicated to astrophysical research and public appreciation of astronomy and operates the DCT in partnership with Boston University, the University of Maryland, the University of Toledo, Northern Arizona University and Yale University. 

The upgrade of the DeVeny optical spectrograph has been funded by a generous grant from John and Ginger Giovale and by a grant from the Mt. Cuba Astronomical Foundation.

Part of the data utilized in this publication were obtained and made available by the The MIT-UH-IRTF Joint Campaign for NEO Reconnaissance. The IRTF is operated by the University of Hawaii under Cooperative Agreement no. NCC 5-538 with the National Aeronautics and Space Administration, Office of Space Science, Planetary Astronomy Program. The MIT component of this work is supported by NASA grant 09-NEOO009-0001, and by the National Science Foundation under Grants Nos. 0506716 and 0907766.

Based on observations obtained at the Gemini Observatory, which is operated by the Association of Universities for Research in Astronomy, Inc., under a cooperative agreement with the NSF on behalf of the Gemini partnership: the National Science Foundation (United States), National Research Council (Canada), CONICYT (Chile), Ministerio de Ciencia, Tecnolog\'{i}a e Innovaci\'{o}n Productiva (Argentina), Minist\'{e}rio da Ci\^{e}ncia, Tecnologia e Inova\c{c}\~{a}o (Brazil), and Korea Astronomy and Space Science Institute (Republic of Korea). The authors wants to acknowledge all the astronomers who obtained the data for this work.

The authors acknowledge support from NASA NEOO grants NNX14AN82G, and NNX17AH06G. D. Polishook is grateful to the Ministry of Science, Technology and Space of the Israeli government for their Ramon fellowship for post-docs.
\appendix

\section{Observing circumstances and taxonomy for the NEOs observed by MANOS}

\begin{deluxetable*}{l|cccccccccccc}
\tablecaption{Observationnal circumstances and spectral analysis results. \label{tab:obs}}
\tablehead{
\colhead{Object} & \colhead{H mag} & \colhead{Obs. Date} & \colhead{V mag} & \colhead{$\Delta$} & \colhead{Airmass} & \colhead{Solar analog} & \colhead{Airmass} & MOID & $\Delta v$ & Phase & Fac. & \colhead{Taxon.} \\
\colhead{} & \colhead{} & \colhead{(YY/MM/DD)} & \colhead{(NEO)} & \colhead{[LD]} & \colhead{(NEO)} & \colhead{(SA)} & \colhead{(SA)} &[AU] & [km/s] & \colhead{[$^\circ$]} & \colhead{} &\colhead{}}
\startdata
1999 SH10 & 22.5 & 14/04/30 & 20.5 & 47.3 & 1.09 & SA107-998 & 1.23 & 0.0094 & 5.5 & 66.8 & GMOSN & V\\
2004 VJ1 & 24.3 & 15/11/04 & 19.6 & 38.6 & 1.07 & SA93-101 & 1.08 & 0.0136 & 5.2 & 3.4 & GMOSN & Cb\\
2007 MK6 & 19.9 & 16/06/15 & 18.5 & 39.6 & 1.12 & SA105-56 & 1.39 & 0.0878 & 15.5 & 89.8 & DCT & O\\
2008 EZ5 & 19.4 & 17/03/19 & 17.7 & 106.7 & 1.23 & SA107-998 & 1.5 & 0.0775 & 6.4 & 8.3 & DCT & Cg\\
2008 HA2 & 24.4 & 15/03/02 & 19.9 & 28.2 & 1.17 & SA102-1081 & 1.15 & 0.0593 & 5.4 & 22.6 & GMOSS & K\\
2009 CP5 & 21.5 & 13/08/16 & 18.1 & 38.3 & 1.07 & SA115-271 & 1.17 & 0.0579 & 5.4 & 35.0 & GMOSN & Sq\\
2010 CE55 & 22.2 & 13/08/11 & 20.9 & 86.8 & 1.06 & SA115-271 & 1.1 & 0.0281 & 4.3 & 42.4 & GMOSN & L\\
2010 CF19 & 21.7 & 13/08/12 & 19.1 & 57.8 & 1.27 & SA115-271 & 1.2 & 0.0327 & 5.5 & 26.9 & GMOSN & Xc\\
2011 BN24 & 20.9 & 13/08/11 & 19.5 & 74.1 & 1.06 & SA115-271 & 1.0 & 0.0156 & 5.5 & 52.4 & GMOSN & X\\
2012 CO46 & 22.9 & 17/09/14 & 19.7 & 64.0 & 1.17 & SA115-271 & 1.09 & 0.0893 & 4.7 & 8.6 & GMOSS & Cgh\\
2013 BO76 & 20.4 & 13/08/12 & 17.2 & 41.2 & 1.2 & SA113-276 & 1.06 & 0.0285 & 6.3 & 34.7 & GMOSN & Q\\
2013 PC7 & 22.4 & 13/08/12 & 19.0 & 56.2 & 1.25 & SA113-276 & 1.32 & 0.1045 & 6.5 & 6.2 & GMOSN & Q\\
2013 PH10 & 23.3 & 13/08/09 & 20.4 & 62.2 & 1.07 & SA112-1333 & 1.19 & 0.1543 & 5.3 & 13.2 & GMOSN & Xe\\
2013 PJ10 & 24.7 & 13/08/11 & 18.6 & 11.3 & 1.11 & SA113-276 & 1.02 & 0.0025 & 5.0 & 41.8 & GMOSN & Sr\\
2013 SR & 24.1 & 13/10/09 & 19.8 & 27.4 & 1.57 & SA115-271 & 1.22 & 0.0695 & 5.3 & 30.4 & GMOSN & Xc\\
2013 VY13 & 21.2 & 13/12/11 & 19.3 & 115.3 & 1.35 & SA98-978 & 1.06 & 0.0422 & 6.3 & 1.3 & GMOSN & S\\
2013 WA44 & 23.6 & 13/12/11 & 20.7 & 68.5 & 1.2 & HD28099 & 1.17 & 0.0206 & 4.2 & 6.1 & GMOSN & Sq\\
2013 WS43 & 22.8 & 13/12/11 & 19.4 & 50.0 & 1.2 & HD28099 & 1.18 & 0.0592 & 5.5 & 19.0 & GMOSN & Sa\\
2013 XV8 & 21.8 & 14/02/28 & 19.0 & 56.7 & 1.47 & SA102-1081 & 1.33 & 0.1044 & 4.9 & 25.3 & GMOSN & X\\
2014 DF80 & 25.9 & 14/03/11 & 20.7 & 30.4 & 1.07 & SA102-1081 & 1.04 & 0.0718 & 5.3 & 4.4 & GMOSN & Ch\\
2014 FA7 & 26.7 & 14/03/27 & 20.1 & 10.7 & 1.29 & SA105-56 & 1.16 & 0.0036 & 5.1 & 23.8 & GMOSN & C\\
2014 FB44 & 25.6 & 14/04/07 & 20.3 & 15.2 & 1.11 & SA105-56 & 1.06 & 0.0161 & 5.5 & 45.9 & GMOSN & L\\
2014 FN33 & 21.1 & 14/04/06 & 19.6 & 96.4 & 1.36 & SA107-998 & 1.4 & 0.1364 & 6.5 & 26.0 & GMOSN & Sq\\
2014 FP47 & 22.3 & 14/04/07 & 19.4 & 65.0 & 1.11 & SA105-56 & 1.32 & 0.0257 & 5.3 & 11.4 & GMOSN & Q\\
2014 GG49 & 25.7 & 14/04/13 & 19.9 & 15.9 & 1.1 & SA107-998 & 1.01 & 0.0077 & 5.8 & 22.7 & GMOSN & Xe\\
2014 HE177 & 25.8 & 14/05/09 & 20.2 & 19.1 & 1.08 & SA105-56 & 1.32 & 0.0479 & 5.9 & 13.5 & GMOSN & C\\
2014 HK129 & 20.9 & 14/05/18 & 20.7 & 133.1 & 1.27 & SA105-56 & 1.14 & 0.0087 & 6.2 & 34.8 & GMOSN & A\\
2014 HS4 & 26.2 & 14/04/29 & 20.4 & 20.5 & 1.22 & SA107-998 & 1.24 & 0.0464 & 5.6 & 9.2 & GMOSN & A\\
2014 HT46 & 26.6 & 14/04/30 & 19.6 & 8.4 & 1.09 & SA107-998 & 1.28 & 0.0127 & 4.3 & 30.4 & GMOSN & Sa\\
2014 JD & 26.3 & 14/05/06 & 19.7 & 11.2 & 1.1 & SA110-361 & 1.1 & 0.0059 & 6.1 & 21.5 & GMOSN & O\\
2014 JJ55 & 25.2 & 14/05/18 & 20.1 & 17.2 & 1.11 & SA107-998 & 1.1 & 0.0196 & 4.9 & 36.3 & GMOSN & Xe\\
2014 MD6 & 21.5 & 14/09/24 & 20.0 & 85.1 & 1.11 & SA93-101 & 1.03 & 0.0721 & 6.2 & 35.6 & GMOSN & Sv\\
2014 OT338 & 21.4 & 14/09/24 & 20.1 & 100.0 & 1.26 & 181-005382 & 1.16 & 0.137 & 6.0 & 25.0 & GMOSN & Sq\\
2014 RC & 26.8 & 14/09/07 & 15.5 & 1.4 & 1.16 & SA113-276 & 1.08 & 0.0005 & 5.8 & 14.5 & GMOSS & Sq\\
2014 RF11 & 23.7 & 14/09/21 & 20.0 & 38.9 & 1.44 & SA112-1333 & 1.07 & 0.0964 & 5.2 & 24.9 & GMOSS & Sr\\
2014 SB145 & 26.3 & 14/10/02 & 18.7 & 7.2 & 1.11 & SA93-101 & 1.07 & 0.0038 & 5.0 & 20.4 & GMOSN & C\\
2014 SF304 & 27.2 & 14/10/03 & 17.8 & 3.2 & 1.07 & SA93-101 & 1.01 & 0.0022 & 4.9 & 15.3 & GMOSN & Q\\
2014 SO142 & 24.4 & 14/10/01 & 20.1 & 38.0 & 1.06 & SA93-101 & 1.06 & 0.0201 & 6.1 & 6.9 & GMOSN & Xc\\
2014 SU1 & 24.8 & 14/10/07 & 20.2 & 15.4 & 1.16 & SA93-101 & 1.21 & 0.0136 & 4.1 & 56.2 & GMOSS & Xe\\
2014 TP57 & 26.4 & 14/10/21 & 18.8 & 8.4 & 1.09 & SA93-101 & 1.06 & 0.0202 & 5.2 & 17.6 & GMOSN & Cb\\
2014 TR57 & 25.2 & 14/10/25 & 20.4 & 22.5 & 1.21 & SA93-101 & 1.09 & 0.0233 & 5.5 & 30.8 & GMOSN & S\\
2014 UC115 & 23.7 & 14/12/18 & 20.4 & 48.9 & 1.41 & HD28099 & 1.06 & 0.1171 & 6.1 & 20.7 & GMOSN & Q\\
2014 UV210 & 27.0 & 14/12/16 & 18.7 & 7.2 & 1.88 & SA98-978 & 1.3 & 0.0183 & 3.9 & 5.5 & GMOSN & Cb\\
2014 VG2 & 22.7 & 14/11/24 & 19.8 & 42.6 & 1.18 & SA93-101 & 1.01 & 0.0146 & 5.1 & 45.4 & GMOSS & Sq\\
2014 WC201 & 26.1 & 14/11/28 & 18.9 & 9.4 & 1.27 & HD28099 & 1.32 & 0.0024 & 6.3 & 13.6 & GMOSN & Sq\\
2014 WE120 & 23.9 & 14/11/28 & 19.8 & 18.8 & 1.22 & SA102-1081 & 1.2 & 0.0285 & 5.0 & 65.1 & GMOSN & O\\
2014 WE121 & 23.5 & 14/12/18 & 20.9 & 46.3 & 1.41 & HD28099 & 1.1 & 0.0763 & 5.0 & 47.3 & GMOSN & Xe\\
2014 WF201 & 25.6 & 14/11/29 & 18.6 & 9.2 & 1.14 & HD28099 & 1.17 & 0.016 & 5.1 & 22.6 & GMOSN & Xc\\
\enddata
\end{deluxetable*}

\begin{deluxetable*}{l|cccccccccccc}
\tablecaption{(continued).}
\addtocounter{table}{-1}
\tablehead{
\colhead{Object} & \colhead{H mag} & \colhead{Obs. Date} & \colhead{V mag} & \colhead{$\Delta$} & \colhead{Airmass} & \colhead{Solar analog} & \colhead{Airmass} & MOID & $\Delta v$ & Phase & Fac. & \colhead{Taxon.} \\
\colhead{} & \colhead{} & \colhead{(YY/MM/DD)} & \colhead{(NEO)} & \colhead{[LD]} & \colhead{(NEO)} & \colhead{(SA)} & \colhead{(SA)} &[AU] & [km/s] & \colhead{[$^\circ$]} & \colhead{} &\colhead{}}
\startdata
2014 WO69 & 23.5 & 14/11/24 & 19.9 & 49.4 & 1.26 & HD28099 & 1.14 & 0.0985 & 6.2 & 9.0 & GMOSN & S\\
2014 WP4 & 24.3 & 14/11/28 & 19.3 & 19.6 & 1.07 & SA98-978 & 1.06 & 0.0406 & 6.7 & 33.9 & GMOSN & A\\
2014 WR6 & 25.3 & 14/11/24 & 19.9 & 18.5 & 1.26 & HD28099 & 1.06 & 0.0437 & 5.5 & 23.2 & GMOSN & Xe\\
2014 WS7 & 27.3 & 14/11/22 & 18.8 & 4.5 & 1.19 & SA98-978 & 1.18 & 0.0111 & 6.1 & 26.6 & GMOSS & Sr\\
2014 WX202 & 29.6 & 14/12/01 & 19.7 & 1.0 & 1.13 & SA93-101 & 1.07 & 0.0004 & 3.9 & 88.4 & GMOSN & Q\\
2014 WX4 & 26.4 & 14/11/19 & 19.8 & 10.7 & 1.18 & SA98-978 & 1.03 & 0.0075 & 5.5 & 25.3 & GMOSN & K\\
2014 WY119 & 26.3 & 14/11/26 & 18.1 & 4.4 & 1.18 & SA93-101 & 1.22 & 0.0086 & 5.0 & 39.3 & GMOSS & Sq\\
2014 YD & 24.2 & 14/12/20 & 20.3 & 25.2 & 1.43 & SA102-1081 & 1.07 & 0.0052 & 4.0 & 52.2 & GMOSN & Xc\\
2014 YD42 & 22.3 & 15/01/08 & 19.8 & 95.6 & 1.09 & SA98-978 & 1.0 & 0.0816 & 6.9 & 2.1 & GMOSN & L\\
2014 YN & 25.8 & 14/12/22 & 19.3 & 10.9 & 1.01 & HD28099 & 1.1 & 0.0053 & 5.7 & 32.1 & GMOSN & Xc\\
2014 YT34 & 24.7 & 15/01/08 & 18.4 & 15.4 & 1.09 & SA98-978 & 1.07 & 0.0354 & 5.8 & 11.6 & GMOSN & Xe\\
2014 YZ8 & 23.7 & 15/01/09 & 20.6 & 29.9 & 1.07 & 180-113477 & 1.31 & 0.0577 & 6.0 & 68.5 & GMOSN & L\\
2015 AK1 & 24.3 & 15/01/14 & 18.5 & 13.9 & 1.11 & SA98-978 & 1.31 & 0.0143 & 6.2 & 33.1 & GMOSN & X\\
2015 AZ43 & 23.8 & 15/02/16 & 17.5 & 7.8 & 1.23 & SA105-56 & 1.68 & 0.0014 & 5.7 & 71.3 & GMOSN & T\\
2015 BF511 & 24.6 & 15/02/02 & 20.4 & 27.5 & 1.32 & SA102-1081 & 1.1 & 0.0053 & 4.9 & 28.5 & GMOSS & D\\
2015 BK4 & 24.9 & 15/01/22 & 17.9 & 9.0 & 1.22 & SA102-1081 & 1.2 & 0.0064 & 5.9 & 24.5 & GMOSS & S\\
2015 BM510 & 25.1 & 15/02/02 & 19.5 & 15.2 & 1.16 & SA98-978 & 1.08 & 0.0236 & 4.8 & 33.5 & GMOSS & D\\
2015 CF & 23.6 & 15/02/16 & 19.9 & 41.1 & 1.07 & SA98-978 & 1.27 & 0.0629 & 6.0 & 24.8 & GMOSN & S\\
2015 CQ13 & 25.7 & 15/02/17 & 18.6 & 6.8 & 1.13 & SA105-56 & 1.1 & 0.0056 & 5.7 & 44.2 & GMOSN & Xe\\
2015 CW13 & 22.7 & 15/03/15 & 19.6 & 27.6 & 1.06 & SA98-978 & 1.04 & 0.006 & 4.9 & 70.2 & GMOSN & A\\
2015 CZ12 & 25.3 & 15/02/16 & 20.1 & 27.2 & 1.07 & SA102-1081 & 1.02 & 0.0283 & 5.6 & 3.6 & GMOSN & K\\
2015 DC54 & 26.1 & 15/02/22 & 20.0 & 13.0 & 1.07 & SA105-56 & 1.03 & 0.0076 & 5.8 & 27.2 & GMOSN & Q\\
2015 DK200 & 25.8 & 15/03/03 & 19.4 & 11.0 & 1.17 & SA105-56 & 1.13 & 0.0148 & 5.2 & 38.9 & GMOSN & Xc\\
2015 DO215 & 26.6 & 15/03/03 & 18.2 & 3.1 & 1.4 & SA102-1081 & 1.22 & 0.0074 & 4.7 & 58.6 & GMOSS & A\\
2015 DP53 & 24.3 & 15/02/22 & 19.3 & 23.0 & 1.23 & SA102-1081 & 1.06 & 0.0507 & 5.6 & 17.2 & GMOSN & S\\
2015 DS & 24.8 & 15/02/25 & 19.9 & 19.8 & 1.18 & SA98-978 & 1.17 & 0.0429 & 4.5 & 34.6 & GMOSS & L\\
2015 DS53 & 24.0 & 15/03/02 & 16.4 & 3.7 & 1.17 & SA102-1081 & 1.17 & 0.008 & 7.3 & 65.1 & GMOSS & Xc\\
2015 DU & 26.6 & 15/02/22 & 19.3 & 8.0 & 1.23 & SA102-1081 & 1.17 & 0.0194 & 4.0 & 23.9 & GMOSN & Sq\\
2015 DZ198 & 24.6 & 15/03/03 & 19.6 & 24.5 & 1.17 & SA105-56 & 1.13 & 0.0116 & 6.0 & 16.1 & GMOSN & Sr\\
2015 EE7 & 20.2 & 15/04/12 & 17.9 & 44.3 & 1.25 & SA98-978 & 1.03 & 0.0677 & 9.3 & 65.2 & GMOSS & Sq\\
2015 EF & 26.8 & 15/03/11 & 19.0 & 9.2 & 1.21 & SA102-1081 & 1.23 & 0.0063 & 6.7 & 3.4 & GMOSS & Sq\\
2015 EK & 26.3 & 15/03/12 & 18.4 & 3.8 & 1.22 & SA98-978 & 1.11 & 0.0067 & 5.6 & 59.0 & GMOSS & Q\\
2015 FC & 26.6 & 15/03/24 & 18.8 & 5.1 & 1.07 & SA102-1081 & 1.03 & 0.0051 & 5.1 & 40.9 & GMOSN & Xe\\
2015 FP & 25.1 & 15/03/24 & 18.2 & 11.6 & 1.07 & SA102-1081 & 1.04 & 0.0245 & 5.5 & 7.6 & GMOSN & Xe\\
2015 FW33 & 25.9 & 15/03/23 & 19.3 & 9.6 & 1.14 & SA102-1081 & 1.08 & 0.0205 & 4.7 & 36.0 & GMOSN & Sq\\
2015 FX33 & 25.8 & 15/03/23 & 19.8 & 14.5 & 1.14 & SA102-1081 & 1.07 & 0.0342 & 5.5 & 25.6 & GMOSN & Xc\\
2015 HS11 & 27.1 & 15/04/28 & 19.6 & 7.6 & 1.07 & SA107-998 & 1.06 & 0.0182 & 4.3 & 19.4 & GMOSN & C\\
2015 HV11 & 24.3 & 15/05/12 & 19.1 & 24.2 & 1.36 & SA105-56 & 1.64 & 0.0587 & 6.1 & 15.1 & DCT & V\\
2015 JF & 26.3 & 15/05/12 & 19.7 & 14.0 & 1.54 & SA105-56 & 1.24 & 0.0101 & 5.2 & 7.3 & DCT & Xe\\
2015 JR & 26.3 & 15/05/12 & 18.9 & 5.6 & 1.75 & SA105-56 & 1.24 & 0.0087 & 6.4 & 45.3 & DCT & O\\
2015 JW & 25.8 & 15/05/20 & 19.6 & 18.3 & 1.39 & SA107-998 & 1.11 & 0.0314 & 5.8 & 3.3 & GMOSS & Cg\\
2015 KA & 26.2 & 15/05/19 & 18.8 & 7.7 & 1.09 & SA105-56 & 1.19 & 0.0157 & 6.0 & 20.8 & GMOSN & Sr\\
2015 KE & 26.2 & 15/05/24 & 20.4 & 14.1 & 1.2 & SA105-56 & 1.21 & 0.0043 & 4.5 & 28.3 & GMOSS & S\\
2015 LQ21 & 24.4 & 15/06/20 & 19.0 & 13.8 & 1.25 & SA110-361 & 1.3 & 0.0258 & 6.1 & 42.4 & DCT & D\\
2015 MC & 24.2 & 15/06/20 & 18.7 & 18.6 & 1.63 & SA107-684 & 1.27 & 0.0101 & 5.6 & 20.8 & DCT & Sq\\
2015 NA14 & 22.0 & 15/07/26 & 17.7 & 24.6 & 1.06 & SA112-1333 & 1.0 & 0.0608 & 5.7 & 40.5 & GMOSN & Sq\\
2015 OM21 & 22.5 & 15/07/27 & 19.5 & 41.1 & 1.18 & SA115-271 & 1.07 & 0.078 & 6.5 & 42.6 & GMOSN & Q\\
\enddata
\end{deluxetable*}

\begin{deluxetable*}{l|cccccccccccc}
\tablecaption{(continued).}
\addtocounter{table}{-1}
\tablehead{
\colhead{Object} & \colhead{H mag} & \colhead{Obs. Date} & \colhead{V mag} & \colhead{$\Delta$} & \colhead{Airmass} & \colhead{Solar analog} & \colhead{Airmass} & MOID & $\Delta v$ & Phase & Fac. & \colhead{Taxon.} \\
\colhead{} & \colhead{} & \colhead{(YY/MM/DD)} & \colhead{(NEO)} & \colhead{[LD]} & \colhead{(NEO)} & \colhead{(SA)} & \colhead{(SA)} &[AU] & [km/s] & \colhead{[$^\circ$]} & \colhead{} &\colhead{}}
\startdata
2015 OQ21 & 27.9 & 15/07/23 & 17.8 & 1.9 & 1.15 & SA110-361 & 1.4 & 0.0022 & 8.7 & 32.1 & GMOSN & V\\
2015 PK9 & 23.7 & 15/08/18 & 18.9 & 22.3 & 1.28 & SA112-1333 & 1.19 & 0.0126 & 5.8 & 30.4 & DCT & Cg\\
2015 QB & 24.2 & 15/08/18 & 19.4 & 23.9 & 1.22 & SA112-1333 & 1.26 & 0.0339 & 5.4 & 22.6 & DCT & K\\
2015 SA & 25.3 & 15/09/19 & 18.7 & 8.5 & 1.26 & SA112-1333 & 1.03 & 0.0052 & 8.6 & 42.8 & GMOSN & Q\\
2015 TD144 & 22.5 & 15/10/19 & 17.7 & 13.3 & 1.09 & HD28099 & 1.04 & 0.0013 & 8.8 & 67.8 & GMOSN & Q\\
2015 TL238 & 24.9 & 15/10/22 & 19.2 & 19.3 & 2.1 & SA93-101 & 1.88 & 0.0277 & 7.5 & 13.3 & GMOSN & Sq\\
2015 TM143 & 23.6 & 15/10/20 & 19.8 & 37.8 & 1.25 & SA115-271 & 1.28 & 0.0286 & 5.7 & 23.0 & GMOSN & Cgh\\
2015 TW237 & 23.1 & 15/11/10 & 20.0 & 57.6 & 1.17 & SA93-101 & 1.09 & 0.0926 & 5.3 & 10.5 & GMOSN & Sr\\
2015 TZ143 & 25.9 & 15/10/22 & 18.6 & 4.3 & 1.33 & SA93-101 & 1.12 & 0.0067 & 6.7 & 56.9 & GMOSN & Q\\
2015 TZ237 & 24.3 & 15/10/21 & 20.0 & 28.8 & 1.18 & SA93-101 & 1.08 & 0.0727 & 5.7 & 30.4 & GMOSN & T\\
2015 VA106 & 22.7 & 15/11/18 & 18.4 & 31.3 & 1.63 & SA93-101 & 1.12 & 0.0106 & 6.2 & 20.4 & GMOSN & Q\\
2015 VE66 & 24.1 & 15/11/16 & 18.0 & 11.6 & 1.65 & SA93-101 & 1.63 & 0.0189 & 6.0 & 35.2 & GMOSN & Sv\\
2015 VG105 & 24.0 & 15/11/15 & 17.9 & 17.0 & 1.16 & SA93-101 & 1.0 & 0.043 & 5.8 & 8.4 & GMOSN & C\\
2015 VN105 & 27.6 & 15/11/15 & 19.9 & 6.0 & 1.16 & SA93-101 & 1.15 & 0.0136 & 6.3 & 32.4 & GMOSN & Xk\\
2015 VO105 & 24.0 & 15/11/16 & 19.7 & 20.6 & 1.65 & SA93-101 & 1.2 & 0.0011 & 4.8 & 54.0 & GMOSN & Sr\\
2015 VO142 & 29.0 & 15/11/21 & 18.9 & 2.1 & 1.63 & SA93-101 & 1.75 & 0.0026 & 4.3 & 30.8 & GMOSN & Sq\\
2015 WA13 & 26.3 & 15/12/05 & 18.8 & 7.6 & 1.09 & HD28099 & 1.27 & 0.0177 & 4.6 & 28.7 & GMOSN & L\\
2015 XB & 24.1 & 15/12/05 & 18.7 & 17.4 & 1.45 & SA93-101 & 1.09 & 0.0362 & 8.0 & 31.0 & GMOSN & Xe\\
2015 XE & 26.2 & 15/12/07 & 19.2 & 19.7 & 1.06 & SA98-978 & 1.0 & 0.0109 & 4.5 & 18.1 & GMOSN & K\\
2015 XM128 & 24.0 & 15/12/28 & 20.0 & 23.1 & 1.35 & SA115-271 & 1.15 & 0.0311 & 6.3 & 67.1 & GMOSN & B\\
2015 XO & 26.3 & 15/12/05 & 20.3 & 21.7 & 1.09 & HD28099 & 1.21 & 0.0536 & 5.4 & 4.2 & GMOSN & X\\
2015 YD & 24.1 & 15/12/30 & 19.9 & 38.2 & 1.08 & SA98-978 & 1.0 & 0.0504 & 5.8 & 10.1 & GMOSN & Sr\\
2015 YD1 & 24.4 & 15/12/29 & 19.7 & 28.4 & 1.24 & SA98-978 & 1.27 & 0.0593 & 6.8 & 22.3 & GMOSN & L\\
2015 YE & 23.4 & 15/12/31 & 20.2 & 53.9 & 1.13 & SA98-978 & 1.08 & 0.1017 & 6.7 & 16.6 & GMOSN & Sq\\
2015 YS9 & 25.9 & 15/12/31 & 19.4 & 12.5 & 1.13 & SA98-978 & 1.01 & 0.004 & 5.2 & 14.1 & GMOSN & Xe\\
2016 BB15 & 24.4 & 16/02/03 & 20.4 & 45.5 & 1.07 & SA102-1081 & 1.04 & 0.0206 & 6.0 & 6.4 & GMOSN & Sv\\
2016 BC15 & 24.9 & 16/02/04 & 20.4 & 30.8 & 1.28 & SA98-978 & 1.01 & 0.0748 & 6.4 & 12.3 & GMOSN & Sq\\
2016 BJ15 & 23.5 & 16/02/03 & 18.5 & 23.4 & 1.07 & SA102-1081 & 1.04 & 0.0163 & 6.1 & 28.0 & GMOSN & Q\\
2016 BW14 & 25.8 & 16/02/04 & 20.4 & 19.2 & 1.28 & SA98-978 & 1.12 & 0.0415 & 6.6 & 23.8 & GMOSN & Xc\\
2016 CF29 & 24.9 & 16/02/07 & 19.9 & 23.0 & 1.18 & SA102-1081 & 1.34 & 0.0308 & 7.2 & 20.9 & GMOSS & Sq\\
2016 CG18 & 28.5 & 16/02/05 & 17.6 & 1.1 & 1.07 & SA102-1081 & 1.15 & 0.0004 & 4.9 & 48.6 & GMOSN & Xe\\
2016 CK29 & 25.6 & 16/02/09 & 20.0 & 18.9 & 1.15 & SA102-1081 & 1.03 & 0.0354 & 5.9 & 11.5 & GMOSN & Sr\\
2016 CL29 & 24.6 & 16/02/07 & 19.6 & 23.3 & 1.08 & SA98-978 & 1.07 & 0.0324 & 7.8 & 22.1 & GMOSN & Q\\
2016 CO29 & 24.9 & 16/02/07 & 20.2 & 25.7 & 1.08 & SA98-978 & 1.08 & 0.022 & 5.5 & 19.2 & GMOSN & L\\
2016 CS247 & 25.8 & 16/02/16 & 18.3 & 8.3 & 1.25 & SA105-56 & 1.14 & 0.0117 & 4.5 & 24.3 & DCT & S\\
2016 CU29 & 26.4 & 16/02/10 & 19.7 & 8.1 & 1.06 & SA102-1081 & 1.07 & 0.0131 & 6.2 & 33.0 & GMOSN & R\\
2016 EB1 & 25.2 & 16/03/06 & 18.8 & 14.5 & 1.45 & SA102-1081 & 1.52 & 0.0134 & 7.1 & 8.7 & GMOSS & Xc\\
2016 EB28 & 23.2 & 16/03/09 & 20.2 & 53.6 & 1.33 & SA105-56 & 1.11 & 0.0039 & 6.6 & 15.9 & GMOSS & Sq\\
2016 EL157 & 26.9 & 16/03/16 & 19.0 & 7.0 & 1.38 & SA105-56 & 1.1 & 0.0074 & 6.3 & 9.0 & GMOSN & Sq\\
2016 EN156 & 27.7 & 16/03/16 & 19.2 & 5.5 & 1.38 & SA105-56 & 1.15 & 0.0019 & 5.0 & 8.6 & GMOSN & K\\
2016 EQ1 & 25.9 & 16/03/10 & 20.5 & 20.0 & 1.11 & SA102-1081 & 1.04 & 0.0053 & 4.7 & 8.5 & GMOSN & T\\
2016 ES1 & 23.7 & 16/03/08 & 20.4 & 35.5 & 1.12 & SA102-1081 & 1.1 & 0.0894 & 6.7 & 28.4 & GMOSN & X\\
2016 FC & 26.5 & 16/03/21 & 20.0 & 11.0 & 1.14 & SA105-56 & 1.06 & 0.0001 & 4.8 & 23.4 & GMOSS & Xc\\
2016 FL12 & 26.1 & 16/04/04 & 20.1 & 15.3 & 1.1 & SA105-56 & 1.21 & 0.0243 & 4.7 & 12.8 & GMOSN & Cb\\
2016 FW13 & 28.6 & 16/04/04 & 19.2 & 1.9 & 1.09 & SA102-1081 & 1.04 & 0.0014 & 5.7 & 27.9 & GMOSN & Xk\\
2016 GB222 & 26.3 & 16/04/19 & 19.4 & 6.3 & 1.24 & SA107-998 & 1.08 & 0.0073 & 5.8 & 38.4 & DCT & Q\\
\enddata
\end{deluxetable*}

\begin{deluxetable*}{l|cccccccccccc}
\tablecaption{(continued).}\addtocounter{table}{-1}
\tablehead{
\colhead{Object} & \colhead{H mag} & \colhead{Obs. Date} & \colhead{V mag} & \colhead{$\Delta$} & \colhead{Airmass} & \colhead{Solar analog} & \colhead{Airmass} & MOID & $\Delta v$ & Phase & Fac. & \colhead{Taxon.} \\
\colhead{} & \colhead{} & \colhead{(YY/MM/DD)} & \colhead{(NEO)} & \colhead{[LD]} & \colhead{(NEO)} & \colhead{(SA)} & \colhead{(SA)} &[AU] & [km/s] & \colhead{[$^\circ$]} & \colhead{} &\colhead{}}
\startdata
2016 GV221 & 25.0 & 16/04/19 & 19.5 & 10.1 & 1.45 & SA107-684 & 1.15 & 0.021 & 5.5 & 70.8 & DCT & Q\\
2016 HB & 24.3 & 16/04/19 & 19.2 & 16.3 & 1.49 & SA107-684 & 1.22 & 0.012 & 5.5 & 41.1 & DCT & Sa\\
2016 HN2 & 23.6 & 16/05/12 & 20.5 & 50.0 & 1.08 & SA110-361 & 1.15 & 0.0611 & 6.1 & 28.0 & GMOSN & Xk\\
2016 HQ19 & 23.8 & 16/05/27 & 20.2 & 24.9 & 1.11 & SA110-361 & 1.18 & 0.0605 & 6.4 & 64.6 & GMOSN & B\\
2016 JV & 25.4 & 16/06/08 & 19.9 & 16.4 & 1.16 & SA110-361 & 1.22 & 0.0366 & 5.9 & 45.9 & GMOSS & B\\
2016 LG49 & 22.4 & 16/06/17 & 17.3 & 16.0 & 1.21 & SA107-998 & 1.13 & 0.0334 & 10.5 & 44.1 & DCT & Sq\\
2016 LO48 & 25.4 & 16/06/15 & 19.4 & 12.2 & 1.34 & A110-361 & 1.15 & 0.021 & 5.4 & 36.4 & DCT & X\\
2016 NC1 & 25.3 & 16/07/16 & 18.9 & 8.0 & 1.21 & SA112-1333 & 1.51 & 0.0187 & 5.8 & 52.1 & GMOSS & A\\
2016 ND1 & 25.4 & 16/07/09 & 19.6 & 10.4 & 1.06 & SA115-271 & 1.04 & 0.0147 & 5.4 & 58.2 & GMOSN & Sq\\
2016 NM15 & 27.4 & 16/07/16 & 19.9 & 9.2 & 1.21 & SA110-361 & 1.22 & 0.0111 & 4.2 & 4.8 & GMOSS & B\\
2016 NN15 & 26.7 & 16/07/08 & 19.0 & 4.3 & 1.08 & SA113-276 & 1.26 & 0.009 & 6.2 & 50.8 & GMOSN & Sv\\
2016 NS & 25.3 & 16/07/17 & 17.9 & 9.6 & 1.16 & SA112-1333 & 1.01 & 0.0205 & 4.8 & 6.9 & GMOSS & Xe\\
2016 PX8 & 27.0 & 16/08/12 & 20.0 & 7.6 & 1.21 & SA93-101 & 1.1 & 0.0155 & 6.4 & 38.1 & GMOSS & S\\
2016 QB2 & 24.2 & 16/09/10 & 19.8 & 36.1 & 1.06 & SA115-271 & 1.07 & 0.0801 & 5.0 & 11.2 & GMOSN & Xk\\
2016 QL44 & 25.1 & 16/09/09 & 20.0 & 26.4 & 1.17 & SA115-271 & 1.13 & 0.0091 & 6.7 & 9.2 & GMOSS & Sq\\
2016 QS11 & 25.8 & 16/09/15 & 19.4 & 12.4 & 1.23 & SA115-271 & 1.23 & 0.0249 & 4.6 & 18.9 & DCT & Xc\\
2016 RB1 & 28.2 & 16/09/07 & 15.3 & 0.5 & 1.22 & SA93-101  & 1.15 & 0.0004 & 6.9 & 29.4 & DCT & Xe\\
2016 RD20 & 24.6 & 16/09/27 & 20.3 & 31.9 & 1.06 & SA93-101 & 1.11 & 0.072 & 6.6 & 12.0 & GMOSN & Q\\
2016 RD34 & 27.4 & 16/09/15 & 19.9 & 3.4 & 1.29 & SA93-101 & 1.11 & 0.0036 & 3.9 & 72.6 & DCT & K\\
2016 RF34 & 24.4 & 16/09/15 & 18.5 & 7.7 & 1.05 & Hya 64 & 1.06 & 0.0107 & 5.7 & 71.4 & DCT & Q\\
2016 RJ18 & 23.5 & 16/09/23 & 20.3 & 54.6 & 1.06 & SA93-101 & 1.25 & 0.1091 & 5.9 & 14.1 & GMOSN & A\\
2016 RL20 & 23.6 & 16/09/27 & 19.4 & 19.1 & 1.13 & SA115-271 & 1.7 & 0.027 & 8.5 & 68.1 & GMOSN & L\\
2016 RM20 & 26.2 & 16/09/24 & 20.1 & 7.8 & 1.23 & SA115-271 & 1.71 & 0.0158 & 5.2 & 65.3 & GMOSS & L\\
2016 RT33 & 23.9 & 16/09/15 & 19.7 & 26.6 & 1.23 & SA115-271 & 1.15 & 0.0512 & 8.8 & 36.2 & DCT & Sv\\
2016 RW & 23.2 & 16/09/30 & 19.9 & 28.7 & 1.09 & SA112-1333 & 1.24 & 0.0155 & 5.2 & 64.6 & GMOSN & Q\\
2016 SA2 & 28.1 & 16/09/28 & 19.1 & 3.5 & 1.18 & SA93-101 & 1.06 & 0.0017 & 4.5 & 22.6 & GMOSS & Sq\\
2016 SW1 & 28.5 & 16/09/28 & 19.0 & 1.5 & 1.06 & HD28099 & 1.09 & 0.002 & 4.7 & 74.8 & GMOSN & Xe\\
2016 SZ1 & 26.4 & 16/09/29 & 19.7 & 6.1 & 1.03 & HD28099 & 1.27 & 0.0108 & 6.0 & 66.9 & GMOSN & L\\
2016 TB57 & 26.1 & 16/10/19 & 19.0 & 8.8 & 1.08 & SA98-978 & 1.22 & 0.0 & 4.1 & 22.6 & GMOSN & S\\
2016 TM56 & 26.7 & 16/10/22 & 18.7 & 5.3 & 1.12 & SA93-101 & 1.12 & 0.0097 & 4.2 & 24.0 & GMOSN & Sq\\
2016 XR23 & 25.3 & 16/12/23 & 20.0 & 15.0 & 1.08 & SA93-101 & 1.22 & 0.003 & 5.2 & 40.8 & GMOSN & Sq\\
2016 YC8 & 24.6 & 17/01/03 & 19.9 & 24.3 & 1.09 & SA98-978 & 1.43 & 0.0006 & 5.5 & 26.4 & GMOSN & Sr\\
2016 YH3 & 24.4 & 16/12/29 & 20.1 & 16.1 & 1.36 & SA102-1081 & 1.24 & 0.0373 & 6.8 & 72.7 & GMOSS & R\\
2016 YM3 & 26.9 & 16/12/29 & 19.3 & 4.4 & 1.36 & SA102-1081 & 1.01 & 0.0091 & 5.6 & 64.7 & GMOSS & X\\
2017 AR4 & 24.6 & 17/01/08 & 19.9 & 31.9 & 1.07 & SA98-978 & 1.03 & 0.0354 & 5.4 & 5.0 & GMOSN & Q\\
2017 AS4 & 26.7 & 17/01/08 & 16.8 & 1.7 & 1.19 & SA102-1081 & 1.19 & 0.001 & 7.5 & 49.4 & GMOSN & S\\
2017 AT4 & 26.7 & 17/01/08 & 19.2 & 11.0 & 1.07 & SA98-978 & 1.0 & 0.028 & 4.5 & 2.9 & GMOSN & Xc\\
2017 BK & 24.0 & 17/01/24 & 19.2 & 15.9 & 1.08 & SA105-56 & 1.12 & 0.0355 & 6.0 & 59.7 & GMOSN & Sq\\
2017 BT & 22.2 & 17/01/25 & 19.9 & 56.0 & 1.13 & SA98-978 & 1.36 & 0.073 & 5.5 & 47.9 & GMOSN & A\\
2017 BU & 25.1 & 17/01/25 & 20.3 & 25.7 & 1.13 & SA98-978 & 1.09 & 0.0229 & 5.8 & 20.5 & GMOSN & T\\
2017 BW & 23.3 & 17/01/31 & 18.2 & 19.7 & 1.16 & SA102-1081 & 1.1 & 0.0103 & 4.7 & 23.8 & GMOSS & Cgh\\
2017 CS & 19.3 & 17/06/04 & 14.9 & 14.0 & 1.31 & SA110-361 & 1.12 & 0.0049 & 6.9 & 62.8 & DCT & C\\
2017 EH4 & 23.9 & 17/03/17 & 19.6 & 34.4 & 1.33 & SA105-56 & 1.33 & 0.0644 & 5.7 & 8.1 & DCT & Sa\\
\enddata
\end{deluxetable*}

\begin{deluxetable*}{l|cccccccccccc}
\tablecaption{(continued).}
\addtocounter{table}{-1}
\tablehead{
\colhead{Object} & \colhead{H mag} & \colhead{Obs. Date} & \colhead{V mag} & \colhead{$\Delta$} & \colhead{Airmass} & \colhead{Solar analog} & \colhead{Airmass} & MOID & $\Delta v$ & Phase & Fac. & \colhead{Taxon.} \\
\colhead{} & \colhead{} & \colhead{(YY/MM/DD)} & \colhead{(NEO)} & \colhead{[LD]} & \colhead{(NEO)} & \colhead{(SA)} & \colhead{(SA)} &[AU] & [km/s] & \colhead{[$^\circ$]} & \colhead{} &\colhead{}}
\startdata
2017 FJ & 28.1 & 17/03/19 & 19.4 & 5.1 & 1.27 & SA102-1081 & 1.18 & 0.0038 & 5.5 & 8.3 & DCT & Sq\\
2017 FK & 27.4 & 17/03/19 & 17.9 & 3.2 & 1.27 & SA102-1081 & 1.07 & 0.0045 & 5.6 & 19.0 & DCT & Sv\\
2017 JM2 & 24.1 & 17/05/14 & 17.0 & 5.9 & 1.39 & SA105-56 & 1.23 & 0.0069 & 9.0 & 45.4 & DCT & S\\
2017 QB35 & 29.3 & 17/09/02 & 18.1 & 1.4 & 1.29 & SA93-101 & 1.18 & 0.0025 & 5.5 & 20.4 & GMOSS & Q\\
2017 QG18 & 27.0 & 17/09/01 & 18.5 & 4.4 & 1.17 & SA115-271 & 1.13 & 0.0045 & 5.2 & 26.2 & GMOSS & Q\\
2017 QR35 & 25.2 & 17/09/06 & 19.6 & 12.4 & 1.4 & SA93-101 & 1.14 & 0.0163 & 5.5 & 46.0 & GMOSN & Xc\\
2017 RB & 28.0 & 17/09/05 & 19.1 & 4.0 & 1.47 & SA93-101 & 1.26 & 0.0088 & 4.9 & 21.2 & GMOSS & Q\\
2017 RB16 & 25.6 & 17/09/26 & 19.9 & 4.7 & 1.63 & SA93-101 & 1.3 & 0.0103 & 6.5 & 105.4 & GMOSS & Cg\\
2017 RS2 & 26.3 & 17/09/23 & 19.8 & 12.6 & 1.68 & SA115-271 & 1.17 & 0.0101 & 4.2 & 11.3 & GMOSS & Q\\
2017 RU2 & 25.9 & 17/09/20 & 19.6 & 7.5 & 1.13 & SA115-271 & 1.33 & 0.0019 & 5.9 & 72.9 & GMOSN & Sr\\
2017 RV2 & 26.3 & 17/09/20 & 20.0 & 12.9 & 1.07 & SA93-101 & 1.5 & 0.0003 & 4.8 & 23.7 & GMOSN & Xe\\
2017 VA15 & 25.1 & 17/10/19 & 17.8 & 8.1 & 1.06 & Hya 64 & 1.04 & 0.0182 & 5.7 & 22.9 & DCT & X\\
2017 VC14 & 28.5 & 17/11/17 & 17.5 & 1.9 & 1.09 & HD28099 & 1.05 & 0.001 & 6.3 & 11.1 & GMOSN & Cg\\
2017 VG1 & 24.0 & 17/11/19 & 19.4 & 24.9 & 1.37 & SA93-101 & 1.12 & 0.01 & 5.9 & 22.2 & GMOSS & Xc\\
2017 VR12 & 20.5 & 18/02/25 & 15.8 & 14.5 & 1.29 & SA105-56 & 1.33 & 0.0077 & 5.1 & 65.1 & DCT & V\\
2017 VV12 & 28.0 & 17/11/16 & 19.6 & 2.4 & 1.22 & SA102-1081 & 1.01 & 0.0007 & 5.2 & 78.7 & GMOSN & Sr\\
2017 VY13 & 26.5 & 17/11/23 & 18.1 & 6.6 & 1.16 & Hya 64 & 1.2 & 0.005 & 5.9 & 8.7 & DCT & Xc\\
2017 VZ14 & 25.0 & 17/11/23 & 16.2 & 3.9 & 1.16 & Hya 64 & 1.33 & 0.0069 & 5.9 & 24.4 & DCT & Xc\\
2017 YF7 & 23.5 & 18/01/22 & 19.9 & 41.8 & 1.3 & SA102-1081 & 1.28 & 0.1002 & 5.0 & 18.6 & GMOSS & S\\
2017 YR3 & 25.3 & 18/01/08 & 20.3 & 22.0 & 1.18 & SA98-978 & 1.12 & 0.007 & 5.0 & 26.1 & GMOSS & Xk\\
2017 YW3 & 26.5 & 18/01/08 & 19.8 & 11.1 & 1.18 & SA98-978 & 1.4 & 0.0069 & 4.2 & 21.0 & GMOSS & Xe\\
2018 AF3 & 22.7 & 18/01/23 & 21.2 & 118.4 & 1.16 & SA98-978 & 1.41 & 0.1091 & 5.6 & 5.4 & GMOSS & L\\
2018 BG1 & 25.5 & 18/01/30 & 19.4 & 14.7 & 1.08 & SA102-1081 & 1.17 & 0.0273 & 5.9 & 17.9 & GMOSN & Sq\\
2018 DT & 27.2 & 18/02/25 & 18.0 & 4.3 & 1.29 & SA105-56 & 1.26 & 0.0105 & 4.1 & 10.9 & DCT & Q\\
2018 DY3 & 25.6 & 18/03/10 & 19.1 & 5.5 & 1.28 & SA105-56 & 1.01 & 0.0084 & 5.1 & 78.0 & DCT & D\\
2018 EH & 24.4 & 18/03/10 & 18.6 & 15.3 & 1.48 & SA105-56 & 1.51 & 0.0101 & 6.6 & 25.1 & DCT & Sq\\
\enddata
\caption{Note: $\Delta$ corresponds to the object-observer distance at the moment of the observation expressed in Lunar distance.}
\end{deluxetable*}

\section{Reflectance spectra for the NEOs observed by MANOS}

\begin{figure*}[ht!]
\plotone{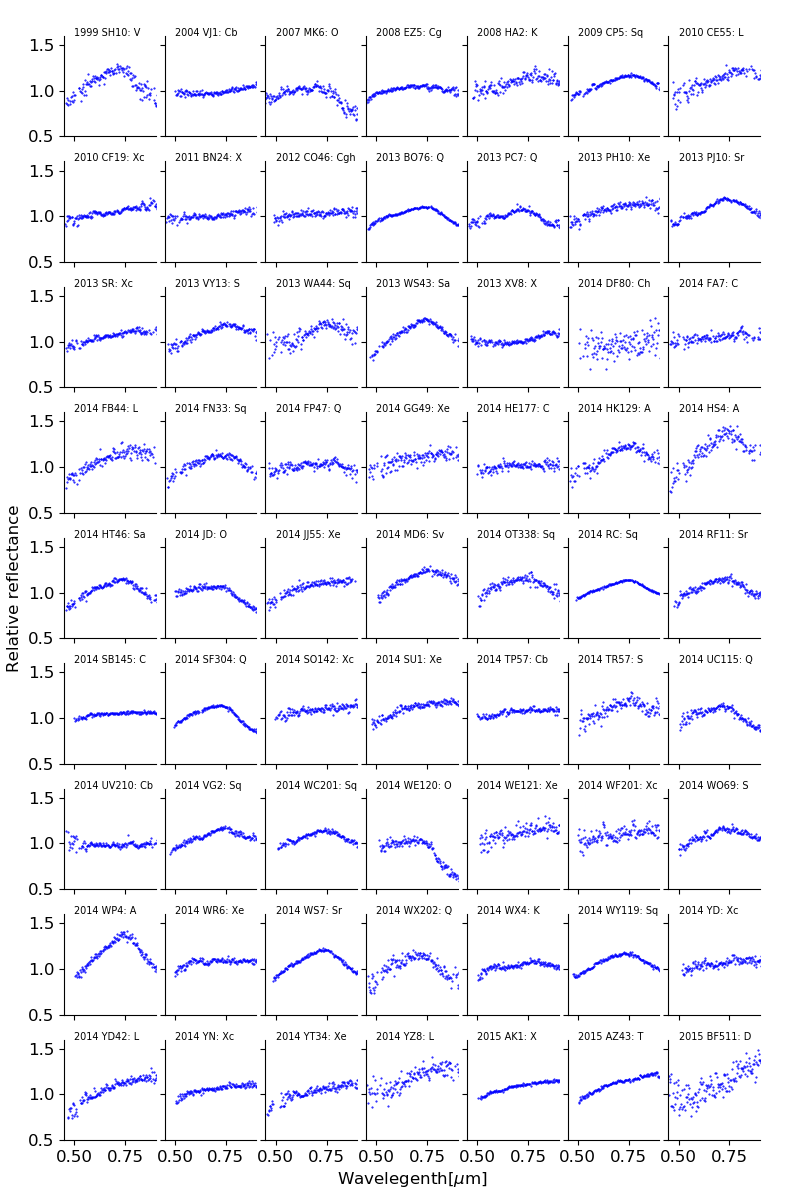}
\caption{Obtained reflectance spectra (normalized at 0.55 $\mu \rm{m}$) obtained in this work. \label{Fig:All}}
\end{figure*}
\begin{figure*}[ht!]
\addtocounter{figure}{-1}
\plotone{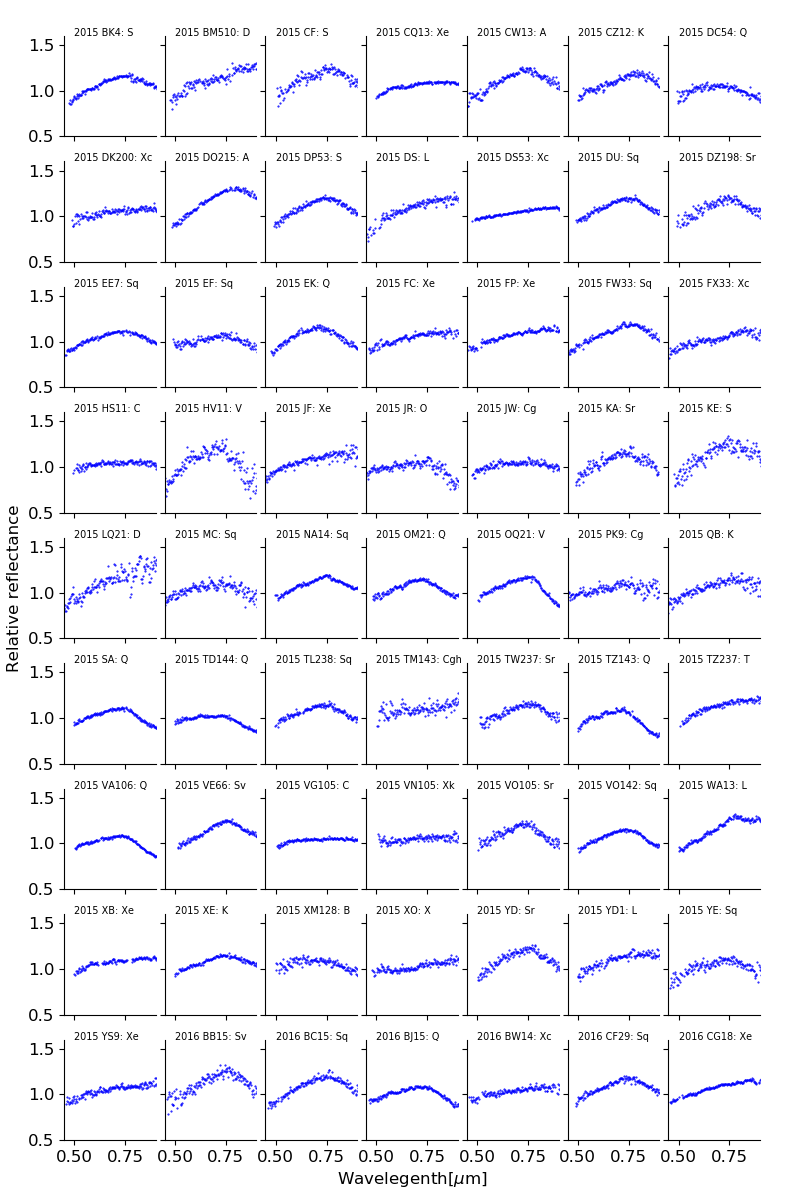}
\caption{(continued).}
\end{figure*}
\begin{figure*}[ht!]
\addtocounter{figure}{-1}
\plotone{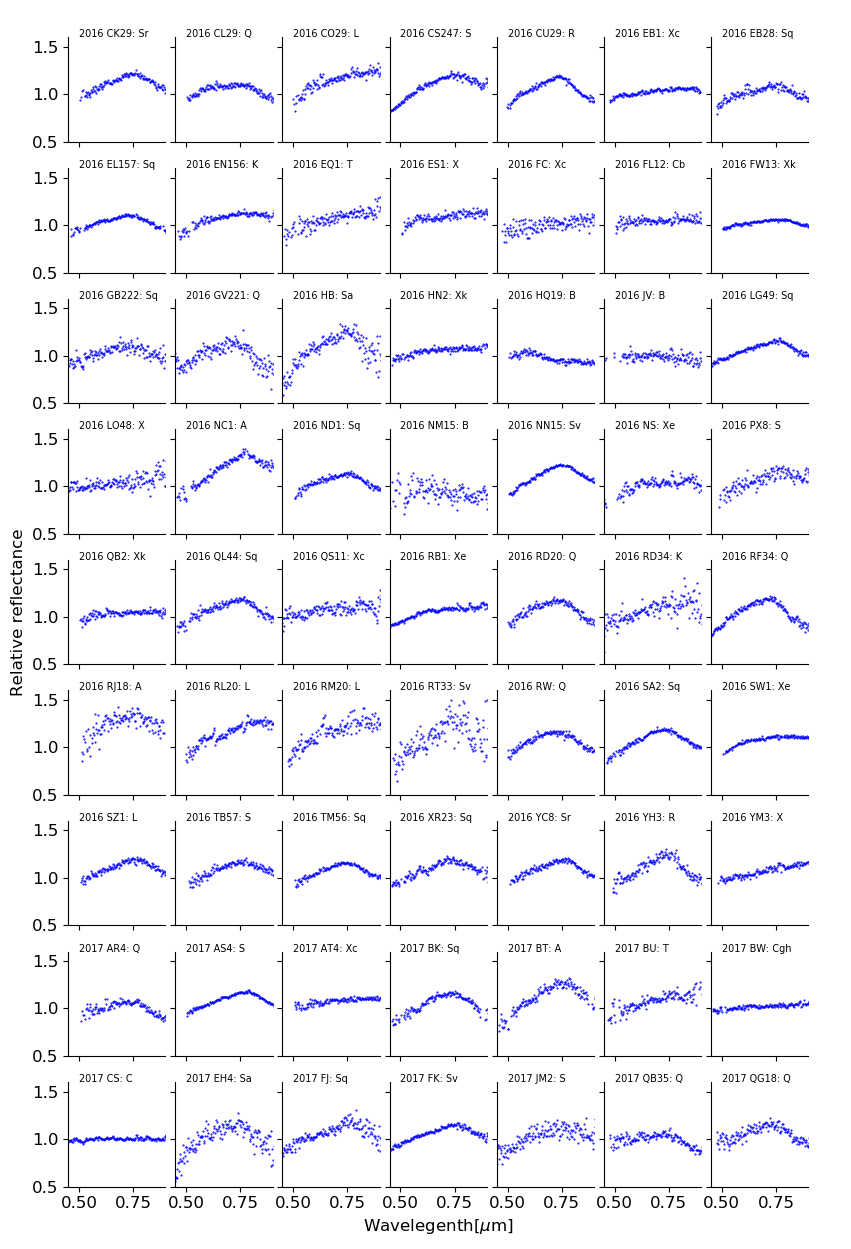}
\caption{(continued).}
\end{figure*}
\begin{figure*}[ht!]
\addtocounter{figure}{-1}
\plotone{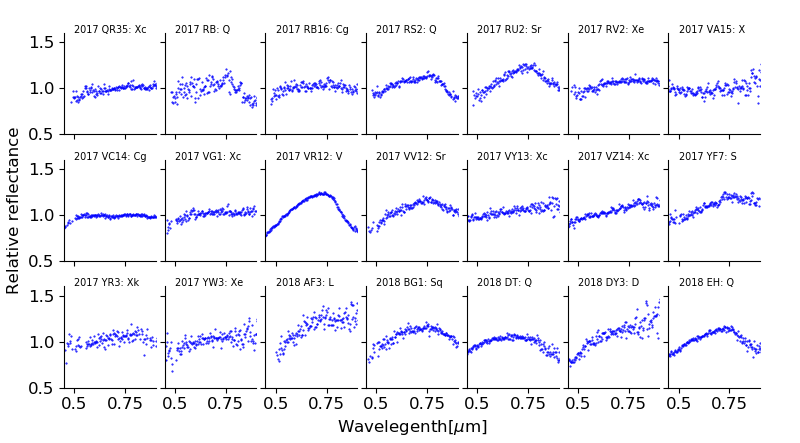}
\caption{(continued).}
\end{figure*}

\bibliography{MANOS1}

\end{document}